\documentclass[aps,prb,twocolumn,longbibliography,superscriptaddress,floatfix,10pt]{revtex4-2}

\usepackage[utf8]{inputenc}
\usepackage{amssymb,amsfonts,amsmath,dsfont}
\usepackage{graphicx}
\usepackage{textcomp}
\usepackage{gensymb}
\usepackage{pifont}
\usepackage{bm}
\usepackage{color}
\usepackage{accents}
\usepackage{mathtools}
\usepackage{nicefrac}
\usepackage{url}
\usepackage{multirow}
\usepackage{hyperref}
\usepackage[capitalize]{cleveref}
\usepackage{tabularx}
\usepackage{upgreek}
\usepackage{chemformula}
\usepackage{xcolor}
\hypersetup{colorlinks=true,citecolor=blue,urlcolor=blue,linkcolor=blue}
\usepackage{cancel}
\usepackage{etoolbox} 
\usepackage{placeins}

\makeatletter
\patchcmd{\subsubsection}{\itshape}{\bfseries\itshape}{}{}
\makeatother

\usepackage{mathrsfs}

\newcommand{\bs}[1]{\boldsymbol{#1}}

\graphicspath{{figures/}}

\makeatletter

\begin{document}
\title{Scattering theory of spin waves by lattice dislocation defects}
\author{Cristobal Larronde}
\affiliation{Departamento de Ciencias F\'isicas, Universidad de La Frontera, Casilla 54-D, Temuco, Chile}
\author{Ignacio Castro}
\affiliation{Departamento de Ciencias F\'isicas, Universidad de La Frontera, Casilla 54-D, Temuco, Chile}
\author{Alvaro S. Nuñez}
\affiliation{Departamento de F{\'i}sica, Facultad de Ciencias Físicas y Matemáticas, CEDENNA, Universidad de Chile, Santiago, Chile}
\author{Roberto E. Troncoso}
\affiliation{Departamento de Física, Facultad de Ciencias, Universidad de Tarapacá, Casilla 7-D, Arica, Chile}
\author{Nicolas Vidal-Silva}
\email{nicolas.vidal@ufrontera.cl}
\affiliation{Departamento de Ciencias F\'isicas, Universidad de La Frontera, Casilla 54-D, Temuco, Chile}

\date{\today}

\begin{abstract}
We investigate spin-wave propagation in magnetic insulators in the presence of lattice dislocations. Within a continuum magnetoelastic framework, we show that the strain fields generated by dislocations induce equilibrium magnetic textures. The morphology of these textures depends sensitively on the dislocation type and acts as a localized scattering potential for spin-wave excitations. As a result, the scattering response exhibits pronounced asymmetries and interference effects governed by the magnetoelastic coupling and the dislocation type. By combining numerical simulations with analytical scattering theory, we compute differential cross sections and frequency-dependent transmission coefficients. Furthermore, analysis of the effective potential landscape reveals that the defect forms a barrier that modulates spin-wave transport and, crucially, breaks the intrinsic reflectionless nature of magnetic domain walls. Our findings identify lattice dislocations as tunable scattering centers, opening new avenues for defect engineering in magnonic devices.
\end{abstract}

\date{\today}

\maketitle

\section{Introduction}
Spin-waves (SW) are collective excitations of magnetically ordered systems \cite{stancil}, which can carry both energy and spin angular momentum.
Their favorable transport properties make them attractive carriers for information processing and storage \cite{chumak}, particularly due to the absence of Joule heating associated with charge currents \cite{flebus,chumak,yuan,chumak-2014}. Consequently, the study of SWs--encompassing the field of magnonics--has emerged as a promising alternative to conventional charge-based electronics \cite{han,stamps}. The discovery of low-damping materials like yttrium iron garnet (YIG) \cite{rezende},
has facilitated the development of functional magnonic devices, such as SW logic gates \cite{khitun,khivintsev}, magnon transistors \cite{chumak-2014}, nonlocal magnon transport devices \cite{cornelissen}, spin–Seebeck generators \cite{adachi}, and coherent magnon–photon coupling platforms \cite{tabuchi}. Despite these advances, SW propagation remains limited by several decay mechanisms. Energy, momentum, and spin are inevitably lost through magnetic damping and interactions with other excitations \cite{chumak2,han}, including magnon-magnon \cite{liu-2023,dzyapko}, magnon-phonons \cite{bozhko,streib,sabiryanov}, and magnon-lattice defects interaction \cite{turski}.

Understanding the SW behavior in real materials requires accounting for the presence of lattice defects, e.g., vacancies, dislocations, or structural disorder. Dislocations are topological defects, characterized by a discontinuity in the lattice and a local breaking of translation symmetry \cite{HirthLothe}. Within continuum elasticity theory, dislocations act as sources of strain and torsion fields, which might couple to various internal degrees of freedom, such as spin or charge \cite{ran,islam}. The role of dislocations in mediating mechanical properties of materials, such as plastic deformation and hardening, is well established in both structural and functional materials \cite{HirthLothe,friedel}. Beyond mechanics, dislocations also influence the transport and response properties, altering the thermal \cite{ninomiya}, electrical \cite{baghani}, and optical \cite{miyajima} properties of the material. Furthermore, the inherent topological nature of dislocations has attracted interest in their role in inducing novel phenomena in superconductivity \cite{teo}, topological materials \cite{saji-2025}, and magnetic insulators \cite{saji}.

A natural coupling between dislocations and spin degrees of freedom arises in systems where magnetism is sensitive to lattice distortions, mediated by magnetoelastic interactions or strain-dependent exchange integrals \cite{kittel}. Previous theoretical works have investigated this interaction, identifying mechanisms for magnon scattering \cite{turski,gestrin,kuchko}, phase shifts \cite{buijnsters}, mode conversion \cite{dobrovolskiy}, localization \cite{zhou,yang-2013}, and topological states \cite{saji,saji-2025}. Theoretical insights from analogous systems, such as electrons in curved space \cite{cortijo} or photons in dislocated photonic crystals \cite{wang-2020,li-light}, further indicate that these defects can induce nontrivial wave phenomena, including scattering behaviors \cite{turski}, wave localization \cite{ran}, and formation of bound states \cite{pereira}. However, these studies have predominantly relied on quantum mechanical formalisms or perturbation theories that treat the dislocation as a local point defect. Consequently, the influence of the strain field generated by a dislocation on the SW properties remains largely unexplored.

In this work, we investigate the interaction of SW and a lattice dislocation within a continuum magnetoelastic framework. Our primary goal is to determine the impact of these structural defects on SW propagation and scattering. To achieve this, we analyze a quasi-two-dimensional system assuming translational invariance along the dislocation line. Furthermore, to obtain an analytical intuition into the scattering mechanism, we employ a one-dimensional approximation. We demonstrate that a dislocation line induces local magnetic textures that act as effective potentials, modulating SW transmission and enabling controlled guiding and filtering. These findings establish a fundamental link between structural defects and magnetic transport, highlighting the potential of dislocation engineering in the design of functional spin-based devices.

This paper is organized as follows. Section \ref{model} formulates the theoretical model, introducing the magnetic free energy and the elastic displacement fields associated with the lattice dislocation. Section \ref{III} characterizes the dislocation-induced magnetic textures, contrasting the analytical insights from a one-dimensional reduction with the full two-dimensional numerical ground states. Section \ref{IV} is dedicated to the scattering dynamics, moving from a one-dimensional calculation of reflection coefficients to a comprehensive numerical analysis of the two-dimensional differential cross-sections. Lastly, Section \ref{conclu} is devoted to conclusions and outlook for further development of the theory in view of outstanding challenges.

\section{Magnetoelastic model}\label{model}
\begin{figure}[t]
    \centering
\includegraphics[width=0.4\textwidth]{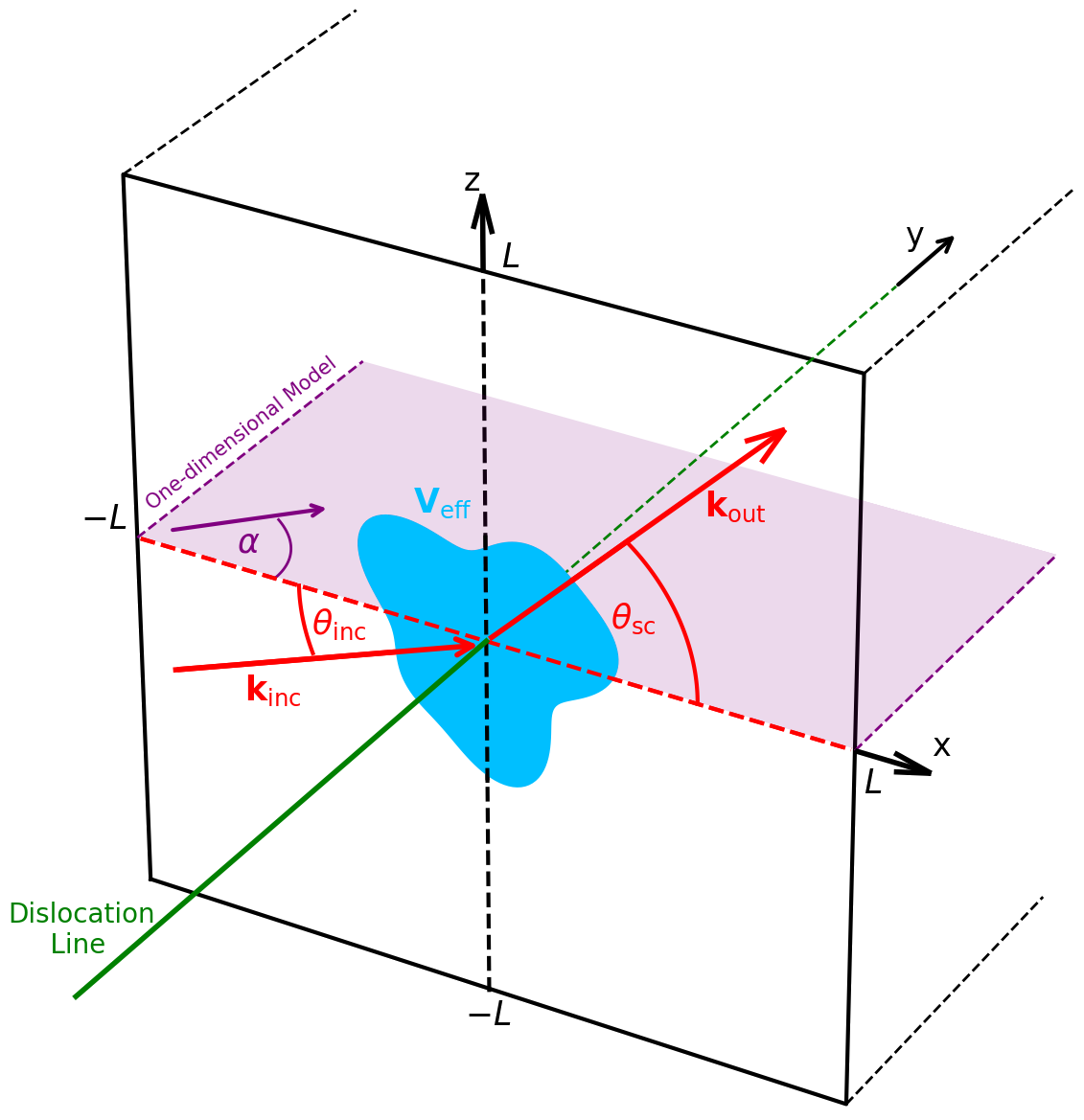}
    \caption{Schematic of SW scattering by an effective potential, $V_{\text{eff}}$, produced by a dislocation line. The vectors $\bm{k}_{\text{inc}}$ and $\bm{k}_{\text{out}}$ denote the incident and scattered SW vectors, respectively. Similarly, $\theta_{\text{inc}}$ and $\theta_{\text{sc}}$ represent the incident and deflection angles within the $xz$-plane, while $\alpha$ defines the azimuthal angle of incidence within the $xy$-plane.}
    \label{fig1}
\end{figure}

We consider a three-dimensional isotropic magnetic insulator with cubic crystal symmetry containing a straight-line dislocation along the $y$-axis, as shown in Fig. \ref{fig1}. The total magnetic and elastic energy density is given by
\begin{equation}
    \mathcal{E} = A\left(\nabla\bm{m}\right)^2 - K_{\text{eff}}\, m_y^2 + b_{1}\sum_{i}m_i^2\epsilon_{ii} + b_2 \sum_{i\neq j} m_im_j\epsilon_{ij},
    \label{energy}
\end{equation}
where $A$ is the exchange stiffness constant and $K_{\text{eff}}$ denotes the effective uniaxial magnetic anisotropy strength along the $y$-axis. This effective anisotropy, $K_{\text{eff}} = K_u + K_{\text{shape}}$, combines the magnetocrystalline anisotropy $K_u$ and the shape anisotropy contribution $K_{\text{shape}} = \mu_0 M_s^2/2$.
The coupling between the magnetization and lattice displacement ${\bs u}$ is described by the magnetoelastic interaction, through the symmetric strain tensor $\epsilon_{ij} = (\partial_i u_j + \partial_j u_i)/2$, with $i,j = x,y,$ and $z$. For isotropic systems \cite{kittel}, the interaction corresponds to the Kittel magnetoelastic energy density with $b_1$ and $b_2$ the magnetoelastic constants. 

Within linear elasticity theory, a straight-line dislocation generates a strain field that modifies the local magnetic energy landscape via the magnetoelastic coupling (MEC). We employ a continuum description of the lattice distortion, treating the magnetic insulator as an isotropic, linear elastic medium. To provide a general description of the scattering phenomena, we consider a mixed straight‑line dislocation along the $y$‑axis, as illustrated in Fig. \ref{fig1}. The corresponding displacement field is decomposed into edge and screw contributions, $\bm{u} = \bm{u}_{\text{edge}} + \bm{u}_{\text{screw}}$, which are defined as \cite{dluzewski}
\begin{align}
    \bm{u}_{\text{edge}} =& \frac{b_x}{2\pi} \left[\left(\arctan{\left(\frac{z}{x}\right)} + \frac{1}{2(1-\nu)}\frac{xz}{x^2+z^2}\right)\hat{\bm{x}} \right.\label{uedge}\\
    &\left.- \left(\frac{1-2\nu}{4(1-\nu)}\ln(x^2 + z^2) + \frac{1}{4(1-\nu)}\frac{x^2-z^2}{x^2+z^2}\right)\hat{\bm{z}}\right],\nonumber\\
    \bm{u}_{\text{screw}} = &\frac{b_y}{2\pi} \arctan\left(\frac{z}{x}\right)\hat{\bm{y}}, \label{uscrew}
\end{align}
where $\nu$ is the Poisson ratio, $b_x$ and $b_y$ correspond to the Burgers vector components. The corresponding strain tensor $\epsilon_{ij}$ is derived from $\bm{u}$ and provided in Appendix \ref{AppendixA}.


\section{Dislocation-Induced Magnetic Textures}\label{III}
In this section, we determine the magnetic ground state in the presence of a dislocation line. In the absence of lattice defects, the unperturbed equilibrium configuration corresponds to a magnetization saturated along the 
$y$-direction. We examine how this uniform state is locally distorted by three representative types of dislocations: mixed, edge, and screw. To assess the role of the MEC strength, we consider two parameter regimes. First, we define a weak-coupling regime using standard magnetoelastic constants for YIG, a magnetic insulator characterized by weak spin–lattice coupling. Second, we investigate a strong-coupling regime, in which the magnetoelastic constants are increased by one order of magnitude. 

We restrict our analysis to a rectangular cross-section of dimensions $2L \times 2L$ in the $xz$-plane, as shown in Fig. \ref{fig1}, and assume the system is sufficiently long in the $y$-direction to ensure translational invariance of the magnetization. Then, parameterizing the magnetization in spherical coordinates, $\bm{m} = (\sin\theta\cos\phi, \sin\theta\sin\phi, \cos\theta)$, with $\theta=\theta(x,z)$ and $\phi=\phi(x,z)$, the equilibrium ground state $\bm{m}_0$ is found by minimizing the total energy density (Eq. \eqref{energy}), leading to the coupled Euler-Lagrange equations
\begin{align}
    \frac{\partial\zeta}{\partial\theta}&=2A\nabla^2\theta -A\left(\nabla\phi\right)^2\sin2\theta + K_{\text{eff}}\,\sin2\theta\sin^2\phi,\label{2dtheta}\\
    \frac{\partial\zeta}{\partial\phi}&=2A\nabla\cdot(\nabla\phi\;\sin^2\theta) +2K_{\text{eff}}\,\sin^2\theta\sin2\phi, \label{2dphi}
\end{align}
where the effects of the dislocation are encoded in the function  $\zeta=\zeta(\theta,\phi)$, defined by 
\begin{align*}
    &\zeta(\theta,\phi) = b_1\sin^2\theta\left[(\epsilon_{xx}-\epsilon_{yy})\cos2\phi + (\epsilon_{xx}+\epsilon_{yy}-2\epsilon_{zz})\right]\nonumber\\
    & \qquad+ 2b_2\big[\epsilon_{xy}\sin^2\theta\sin2\phi + (\epsilon_{xz}\cos\phi + \epsilon_{yz}\sin\phi)\sin2\theta\big].
    \label{MECfunc}
\end{align*}
Dislocation-induced fields break translational invariance and, as a result, reorient the magnetic vector into textured configurations. Because the strain field is spatially localized and decays algebraically away from the dislocation core, the magnetic ground state is confined to a finite region surrounding the defect. Far from the dislocation line, $\bm{m}_0$ asymptotically approaches the uniform defect-free equilibrium state.

\subsection{One-dimensional approximation}\label{IIIA}
To obtain physical insight into dislocation-induced magnetic distortions, we introduce a one-dimensional model defined by the $xy$-plane at $z=0$, as illustrated by the purple plane in Fig. \ref{fig1}. As before, we assume translational invariance along the dislocation line, such that the magnetization varies only along the 
$x$-axis, ${\bs m} = {\bs m}(x)$, and the strain tensor reduces to
$\epsilon_{ij}(x) \equiv \epsilon_{ij}(x,0)$. This description is appropriate for laterally confined geometries, such as narrow magnetic strips, in which transverse modes are suppressed, and the dynamics becomes effectively one-dimensional \cite{flebus}.

For this model, we consider that the effective anisotropy is dominated by a uniaxial term $K_u$ oriented along the $z$-axis (perpendicular to the plane). We assume this crystalline anisotropy is much larger than the effective in-plane contributions, allowing us to approximate $K_{\text{eff}} \approx K_u$. Such a strong perpendicular anisotropy regime is experimentally expected in thin films and engineered heterostructures, where interfacial effects can generate sufficiently strong perpendicular magnetic anisotropy to overcome the shape anisotropy that would otherwise favor in-plane magnetization \cite{masood}. As a result, the magnetic easy-axis reorients perpendicular to the structural plane, allowing the in-plane anisotropy to be neglected. The unperturbed equilibrium state, therefore, corresponds to a uniform magnetization saturated along the $z$-direction.

To analyze the local distortion induced by the dislocation, we enforce the ansatz $\phi(x) = \phi_0$ as a constant and $\theta(x) = \theta_0(x)$. Thus, Eq. (\ref{2dphi}) reduces to $b_x \sin\phi_0 - b_y(1-\nu)\cos\phi_0 = 0$, and $\phi_0  = \arctan\left({(1-\nu)b_y}/{b_x}\right)$. Substituting this equilibrium angle into Eq. (\ref{2dtheta}) yields
\begin{equation}
    \frac{d^2\theta_0}{dx^2} = \frac{K_u}{2A}\sin 2\theta_0 + \frac{\beta}{x}\cos 2\theta_0,
    \label{thetamin}
\end{equation}
where the second term constitutes the dislocation contribution, with $\beta = {b_2}\sqrt{b_x^2 + b_y^2(1 - \nu)^2} / {4\pi A (1-\nu)}$. In particular, when $\beta = 0$, Eq. (\ref{thetamin}) reduces to the well-known problem of magnetic domain walls \cite{lee,kim}. We solve Eq. (\ref{thetamin}) numerically using the Newton-Raphson method, considering two boundary conditions: 1) homogeneous conditions  ($\theta_0(\pm L)=0$) and 2) domain-wall (DW) conditions ($\theta_0(-L)=0$ and $\theta_0(+L)=\pi$). We consider material parameters representative of YIG at low temperatures \cite{klingler,yaqi,nilsen,rabier}: exchange stiffness $A = 3.65\times10^{-12}$ J/m, saturation magnetization $M_s = 1.6\times10^5$ A/m, Burgers vectors $b_x = b_y = 1.256$ nm, and Poisson ratio $\nu = 0.3$. Regarding the anisotropy, we use a single value $K_{\text{eff}} = 2.5\times10^3$ J/m$^3$  for both two-dimensional and one-dimensional models. The magnetoelastic constants are $b_1= 2.3\times10^5$ J/m$^3$ and $b_2= 8\times10^5$ J/m$^3$ (weak regime). For the strong regime, we set $b_1 = 2.3\times10^6$~J/m$^3$ and $b_2 = 8\times10^6$~J/m$^3$.

\begin{figure*}[!htb]
\centering
\includegraphics[width=.8\textwidth]{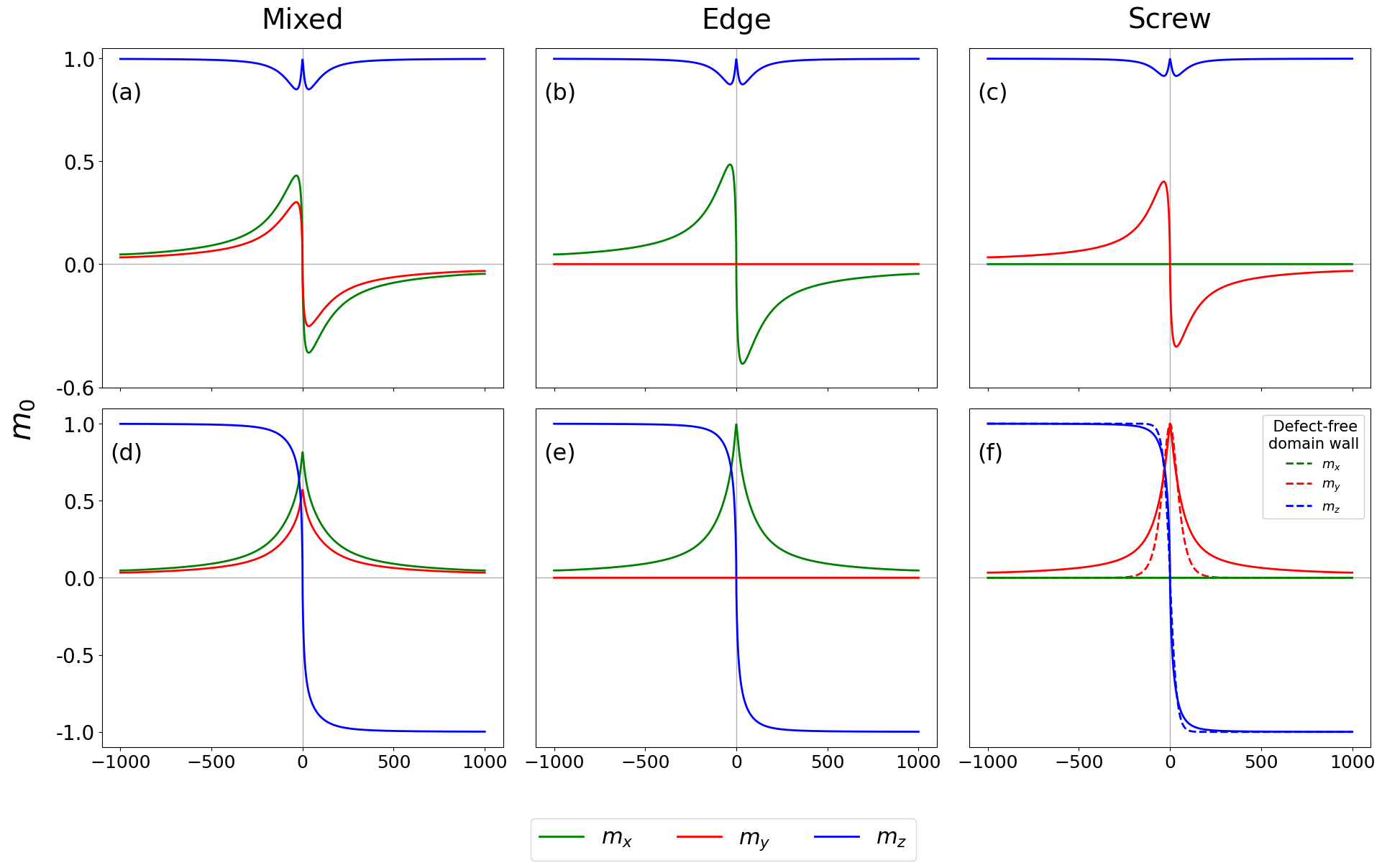}
    \caption{Relaxed magnetization profiles for the one-dimensional model in the weak regime. Panels (a)-(c) show the relaxed profiles for the homogeneous boundary conditions, and panels (d)-(f) show the relaxed profiles for the domain wall conditions. Columns correspond to mixed, edge, and screw dislocation types, respectively. Additionally, panel (f) shows the defect-free Bloch wall profile (dashed lines) \cite{schryer}.}
    \label{fig_GS1}
\end{figure*}
In Fig. \ref{fig_GS1}, we display the magnetization components of the ground states induced by the dislocation line. For the homogeneous conditions in the weak regime, the dislocation induces a smooth deviation of the $z$-component of the magnetization field around the core, independent of the dislocation type, while selectively suppressing either the $m_x$ (edge) or $m_y$ (screw) component (Fig. \ref{fig_GS1}(a)-(c)). Similarly, for the DW conditions, the dislocation type dictates the structure of the resulting domain wall: the edge dislocation induces a Néel-type wall (Fig. \ref{fig_GS1}(e)), while the screw dislocation produces a Bloch-type wall (solid lines in Fig. \ref{fig_GS1}(f)). The mixed dislocation yields a hybrid domain wall structure, where the transition involves out-of-plane and in-plane magnetization, as illustrated in Fig. \ref{fig_GS1}(d). Comparing these profiles to the defect-free Bloch wall (dashed lines in Fig. \ref{fig_GS1}(f)) reveals that the MEC broadens the wall width, indicating a weak pinning effect. This selective dislocation-driven formation of Néel or Bloch walls could enable the identification of dislocation types via magnetic imaging or the tailored design of domain wall types in strained magnonic devices. In the strong regime, the magnetoelastic contribution dominates over exchange near the dislocation core, leading to stronger distortions of the configurations shown in Fig. \ref{fig_GS1}. Further details are provided in Appendix \ref{AppendixB}.
\begin{figure}[t]
    \centering
    \includegraphics[width=0.5\textwidth]{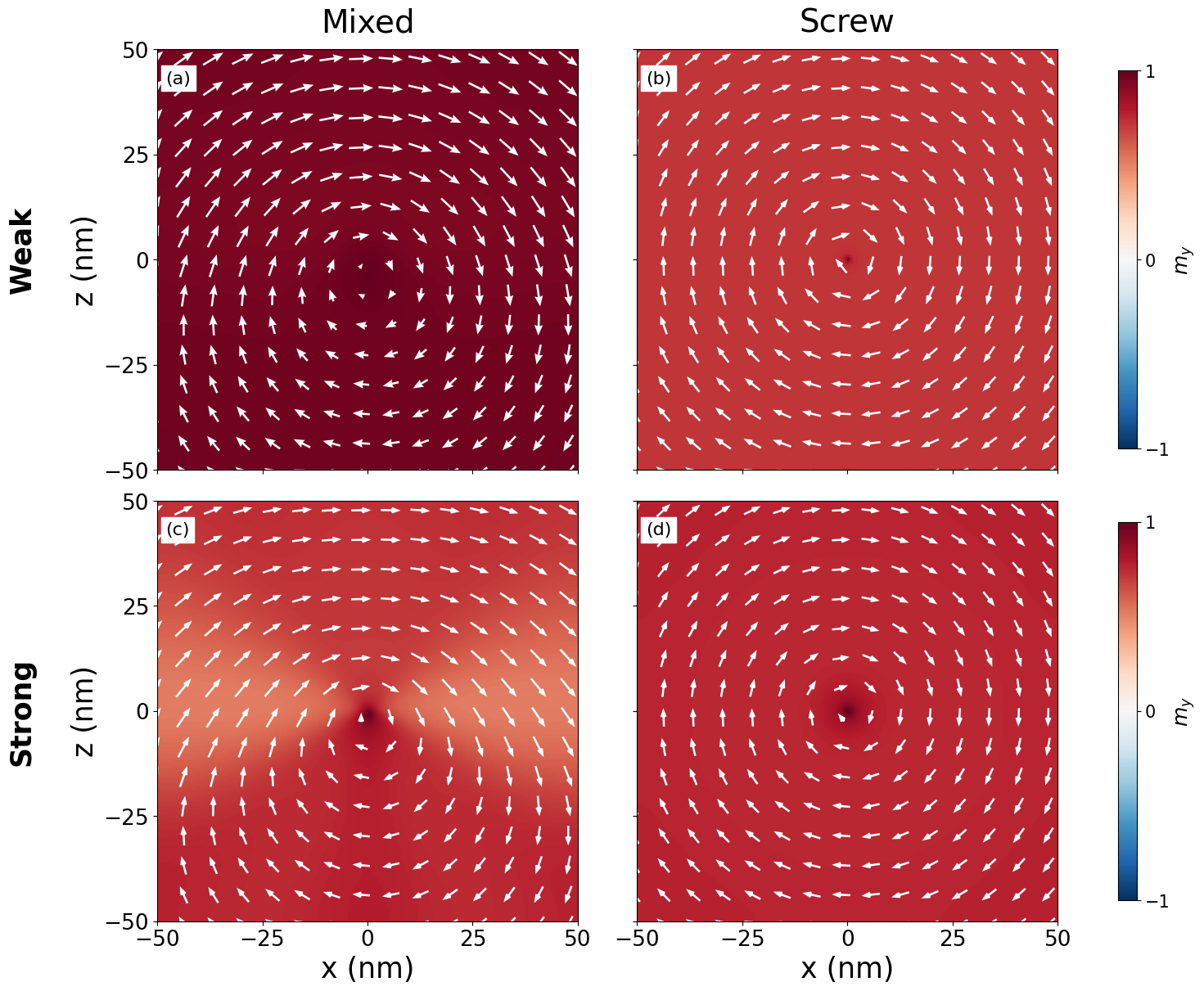}
    \caption{Magnetic ground states in the presence of different dislocation types. The configurations were obtained from an initial ferromagnetic state along the $y$-axis for a (a) mixed and (b) screw dislocation. For comparison, panels (c) and (d) show the strong coupling regime for a mixed and screw dislocation, respectively. Arrows represent the in-plane magnetization components $(m_x, m_z)$, while the color scale encodes the out-of-plane component $m_y$.}
    \label{fig2}
\end{figure}

\subsection{Magnetic textures in the 3D bulk}\label{IIIB}
We now study the magnetic ground state in the full three-dimensional system.
We numerically solve the Landau--Lifshitz--Gilbert (LLG) equation, 
\begin{align}
    \dot{\bm{m}} = \gamma\mu_0(\bm{m}\times\bm{H}_{\text{eff}} + \alpha_G\; \bm{m}\times\dot{\bm{m}}),
    \label{LLG}
\end{align}
where $\gamma$ is the gyromagnetic ratio, $\alpha_G$ is the Gilbert damping constant, $\mu_0$ is the vacuum permeability, and the effective field is $\bm{H}_{\text{eff}} = -{1}/({\mu_0 M_s})\;{\delta\mathcal{E}}/{\delta\bm{m}}$, with $M_s$ the saturation magnetization. This approach automatically enforces the constraint $|{\bs m}| =1$, handles the strong nonlinearity of the coupled magnetoelastic fields, and captures the full precessional and dissipative response of the spin system. We integrate Eq. (\ref{LLG}) from a saturated initial condition using a high damping parameter ($\alpha_G = 0.5$) to ensure rapid convergence. The integration was performed using a fourth-order Runge–Kutta (RK4) method with a time step of $\Delta t = 10^{-15}$ s. The simulation domain spans $200\times200$ nm$^2$, discretized on a $501\times501$ grid, with free boundary conditions applied on all edges. Convergence is considered achieved when the maximum torque satisfies $|\bm{m}\times\bm{H}_{\mathrm{eff}}|<10^{-3}$ A$^2$/m$^2$.

Figure \ref{fig2} displays the magnetic equilibrium configurations for the mixed and screw dislocations for the different regimes. While the total simulated system extends from -100 to 100 nm, we present a magnified view of the central region (from -50 to 50 nm) to highlight the structure of the magnetic textures. The dislocation line modifies the saturated state, creating a local distortion that depends critically on the type of dislocation. 

In the weak regime, the mixed dislocation induces a vortex-like rotation in the $xz$-plane, where the in-plane magnetization components curl smoothly while the out-of-plane component $m_y$ remains near saturation (Fig. \ref{fig2}(a)). In contrast, the screw dislocation generates a tighter in-plane curl with a less pronounced out-of-plane magnetization, as shown in Fig. \ref{fig2}(b), indicating stronger suppression of $m_y$ and larger $m_x$ and $m_z$. In the strong regime, the mixed dislocation texture strongly suppresses $m_y$ along the $z=0$ plane, visible as distinct white regions at Fig. \ref{fig2}(c). This indicates a complete reorientation of the magnetization, favoring the formation of a magnetic vortex. The screw dislocation in the strong regime maintains the in-plane curl, partially recovering $m_y$ as the colormap becomes darker (Fig. \ref{fig2}(d)). 

In all cases shown, the induced textures are confined to the vicinity of the dislocation core. At large distances, the magnetoelastic coupling vanishes, and the magnetization asymptotically recovers the uniform saturated state. These results are consistent with previous observations \cite{berbil}, where a lattice dislocation acts as a nucleation center for magnetic vortices. Finally, the texture generated by the edge dislocation is not shown as the relaxed magnetization remains essentially unchanged in both coupling regimes. Since the magnetization is initially saturated along the $y$-direction, the magnetoelastic energy density associated with the edge dislocation strain field vanishes. As a result, the elastic deformation does not modify the magnetic energy landscape, and the uniform magnetic configuration remains unaltered.

\section{Spin Wave Dynamics and Scattering}\label{IV}
In this section, we analyze the propagation and scattering of SWs across the region with the dislocation line. We consider the linear SW dynamics around the equilibrium magnetization previously found, and assume a low-temperature regime where the propagation of elastic waves is negligible. To describe the excitations, we substitute the ansatz ${\bs m} = {\bs m}_0+\delta{\bs m}$ into the LLG equation, where ${\bs m}_0$ corresponds to the magnetic ground state found in the past section, and $\delta{\bs m}$ stands for the SW contribution. Since the magnetic ground state is textured, we project the resulting magnetization fluctuations, $\delta{\bs m}$, onto a local orthonormal basis $(\hat{\bm{e}}_1, \hat{\bm{e}}_2)$, such that the quantization axis is along $\bm{m}_0$, and thus $\bm{m}_0=\hat{\bm{e}}_1\times\hat{\bm{e}}_2$. The equation of motion for the SWs is given by 
\begin{equation}
\frac{\partial{\psi({\bs r},t)}}{\partial t} = (\mathcal{H}_0+V({\bs r})){\psi({\bs r},t)},
    \label{schro-like}
\end{equation}
where the field $\psi = (\delta m_1, \delta m_2)^T$, $\mathcal{H}_0$ represents the Hamiltonian for SWs in perfect lattices, and $V$ describes the scattering potential induced by the dislocation, which are detailed in Appendix \ref{AppendixA}. 

\subsection{One-dimensional scattering}
We first study the scattering problem between SWs and the dislocation line in the one-dimensional model, addressed in Section \ref{IIIA}.  To describe the fluctuations, we consider the orthonormal basis where $\hat{\bm{e}}_1 = \hat{\bm{e}}_2\times{\bs m}_0$ and $\hat{\bm{e}}_2 = (-\sin\phi, \cos\phi, 0)$. Here, $\theta = \theta_0(x)$ and $\phi = \phi_0$ are the angles defining the equilibrium magnetization. In this basis, and assuming harmonic solutions, $\psi(x,y,t) = \psi(x) e^{i(k_yy-\omega t)}$, Eq. (\ref{schro-like}) reduces to the one-dimensional Schrödinger-like equation
\begin{equation}
\frac{\partial^2 \psi}{\partial x^2} + \frac{M_s}{2A\gamma}\left[\omega - \gamma\left( V_{\text{eff}}(x) +\frac{2 A}{M_s}k_y^2\right)\right]\psi(x)= 0,
\label{schrodinger}
\end{equation}
where we have defined
\begin{align}
    V_{\text{eff}}(x) =& \frac{2K_u}{M_S}\cos2\theta_0(x) - \frac{4A\beta}{xM_S}\sin2\theta_0(x),\label{psipotential}\\
    \beta =& \frac{b_2}{4\pi A (1-\nu)}\sqrt{b_x^2 + b_y^2(1 - \nu)^2}\label{betaconst}.
\end{align}
In the following, we focus on the resulting potential and the scattering coefficients for the two distinct equilibrium magnetizations: (1) the quasi-homogeneous case, obtained under the homogeneous boundary conditions, and (2) the domain wall case, stabilized by domain-wall boundary conditions.

\subsubsection{Quasi-homogeneous case}\label{homocase}
\begin{figure}
\centering
\includegraphics[width=.48\textwidth]{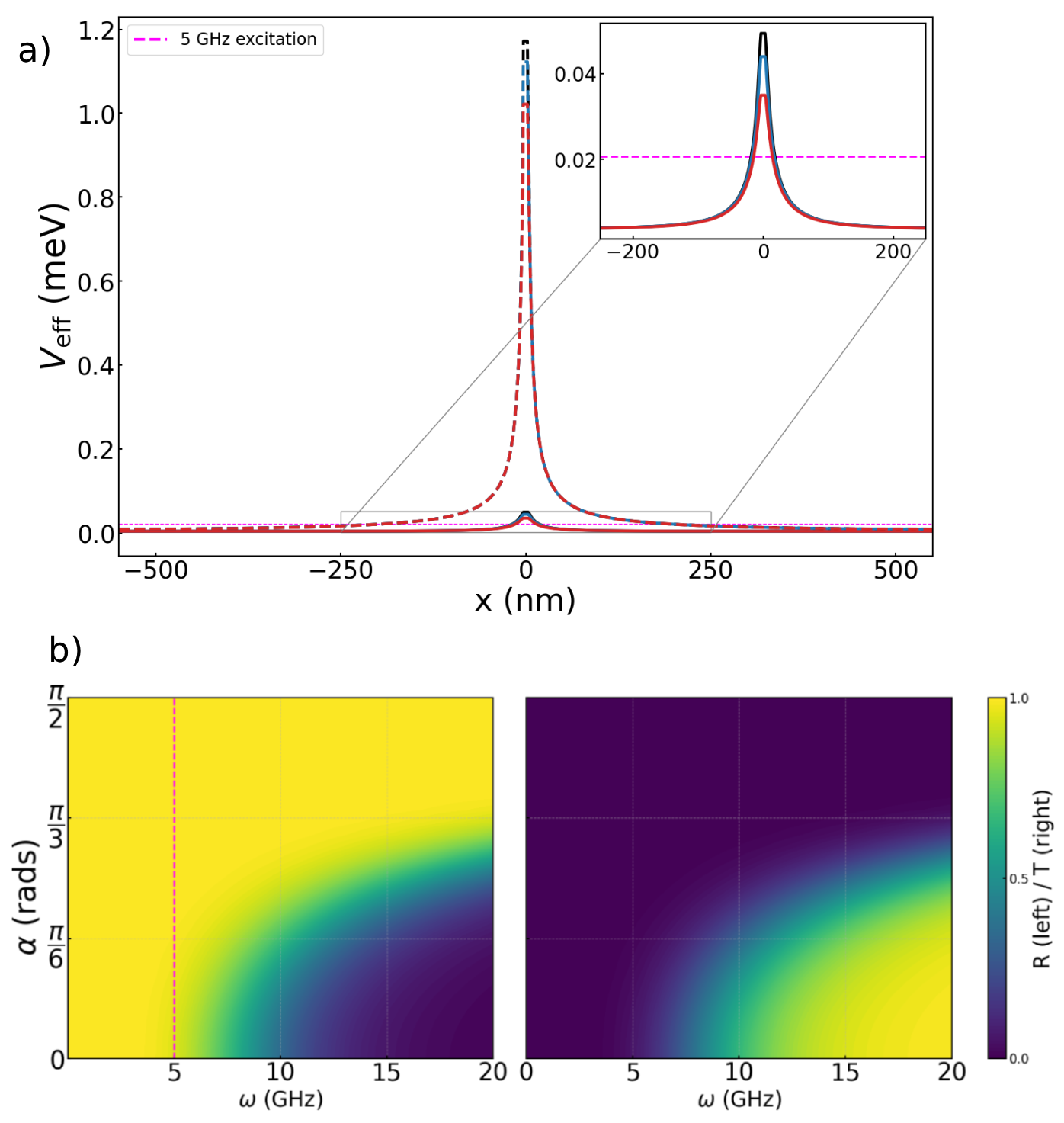}
\caption{(a) One-dimensional effective potential profiles for different dislocation types in the quasi-homogeneous case. The inset provides a magnified view of the weak coupling regime (solid lines) near the origin. Dashed lines represent the strong coupling regime, and the horizontal pink line indicates the 5 GHz excitation level. (b) Reflection (left) and transmission (right) coefficients for SW scattering as a function of frequency $\omega$ and incident angle $\alpha$.}
\label{potentialxRT}
\end{figure}

Figure \ref{potentialxRT}(a) shows the effective potential for each dislocation type in both coupling regimes. We focus on normal incidence without loss of generality ($\alpha = 0$ in Fig. \ref{fig1}), as a finite $k_y$ acts as a constant energy shift that rescales the potential barrier. In the strong-coupling regime, the effective potential barrier reaches $\sim 1.2$ meV, exceeding typical SW excitation energies in magnetic insulators, $\hbar\omega\sim 10-100 \mu$eV \cite{kruglyak,barman}. Consequently, the incident SW satisfies $\hbar\omega \ll V_{\text{eff}}$, resulting in an exponentially suppressed transmission and causing the dislocation to effectively act as a hard-wall scatterer. By contrast, in the weak-coupling regime, the barrier height is only $\sim$40 $\mu$eV, comparable to typical SW energies, such that tunneling and partial transmission are expected. The remaining differences between dislocation types arise from the dependence of the effective potential on the Burgers-vector components entering Eq. (\ref{betaconst}).

The reflection (R) and transmission (T) coefficients for a mixed dislocation in the weak regime are shown in Fig. \ref{potentialxRT}(b). Since the effective potential for the mixed, edge, and screw dislocation has comparable magnitudes (see Fig. \ref{potentialxRT}(a)), the scattering profiles shown here are representative of all dislocation types. At low frequencies ($\omega\lesssim5$ GHz, depicted by the vertical dashed line in Fig \ref{potentialxRT}(b)), the SW energy are insufficient to overcome the potential barrier, resulting in a reflection-dominated region across all incident angles $\alpha$ (Fig. \ref{potentialxRT}(b) left panel). As frequency increases, we observe a smooth crossover to a high-transmission region for $\alpha\lesssim\pi/3$, as shown in Fig. \ref{potentialxRT}(b) right panel. For the strong regime, the system exhibits similar behavior, but the transmission threshold shifts to significantly higher frequencies due to the deeper potential barrier, as shown in Fig. \ref{potentialxRT_strong} in Appendix \ref{AppendixB}.

\subsubsection{Domain-wall case}\label{DWcase}
\begin{figure}
    \centering
    \includegraphics[width=0.48\textwidth]{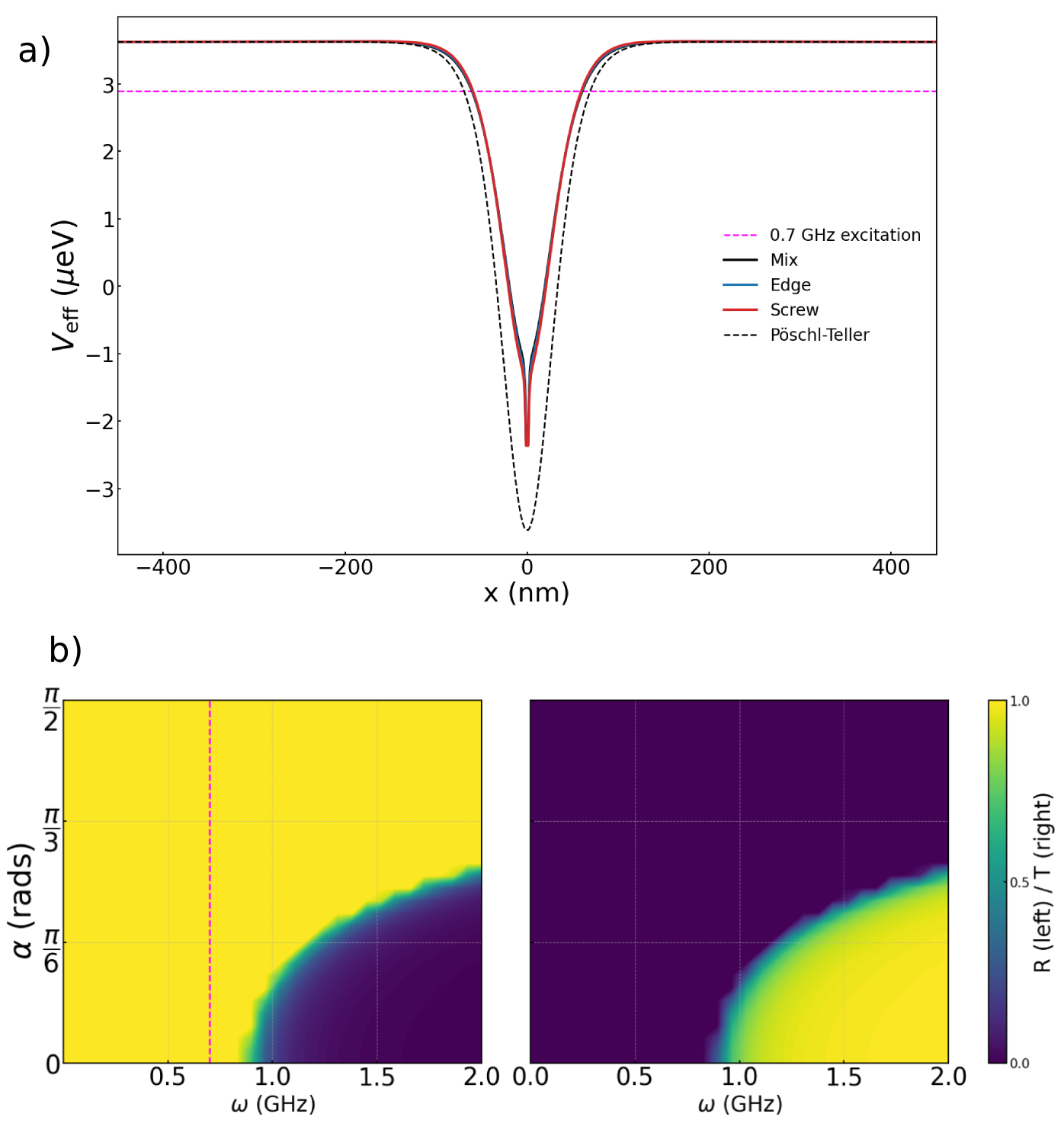}
    \caption{(a) One-dimensional effective potential for the $b_{1,2} = 8\times10^4$ J/m$^3$ limit. Profiles for mixed (black), edge (blue), and screw (red) dislocations are compared to the analytical Pöschl-Teller potential (dashed black line). The horizontal pink line indicates the 0.7 GHz excitation level. (b) Reflection (left) and transmission (right) coefficients for SW scattering as a function of frequency $\omega$ and incident angle $\alpha$.}
    \label{weakpotentialxRT}
\end{figure}
We now consider SW scattering in the domain wall case. For the weak and strong regimes, the dislocation potential completely dominates the scattering behavior, producing a repulsive barrier. Consequently, the resulting reflection and transmission coefficients (Fig. \ref{potentialxRT_DW} in Appendix \ref{AppendixB}) are identical to those obtained for the quasi-homogeneous case (Fig. \ref{potentialxRT}(b)). This indicates that the dislocation line masks the intrinsic domain-wall characteristics, rendering the scattering problem equivalent to that of a defect in a uniformly magnetized background.

A qualitatively distinct regime emerges when the MEC is reduced to a value one order of magnitude below our weak regime. In this ultra-weak coupling limit, the magnetoelastic energy is sufficiently suppressed that the intrinsic nature of the domain wall prevails. As shown in Fig. \ref{weakpotentialxRT}(a), the effective potential for all dislocation types (solid lines) converges to the attractive Pöschl-Teller profile characteristic of a pure domain wall (dashed line). Here, the excitation energy (pink line at 0.7 GHz) interacts with the shallow well rather than a rigid barrier. Consequently, the scattering coefficients (Fig. \ref{weakpotentialxRT}(b)) recover the standard behavior of domain wall propagation \cite{thiele}, where the potential well allows only transmission for frequencies above 1 GHz and small angles, while reflecting lower frequencies due to the finite depth of the potential well. Thus, the MEC strength acts as a critical control parameter, tuning the system from a regime governed by intrinsic domain wall dynamics (ultralow limit) to one dominated entirely by dislocation potential.
\begin{figure*}[t]
    \centering
    \includegraphics[width=0.8\textwidth]{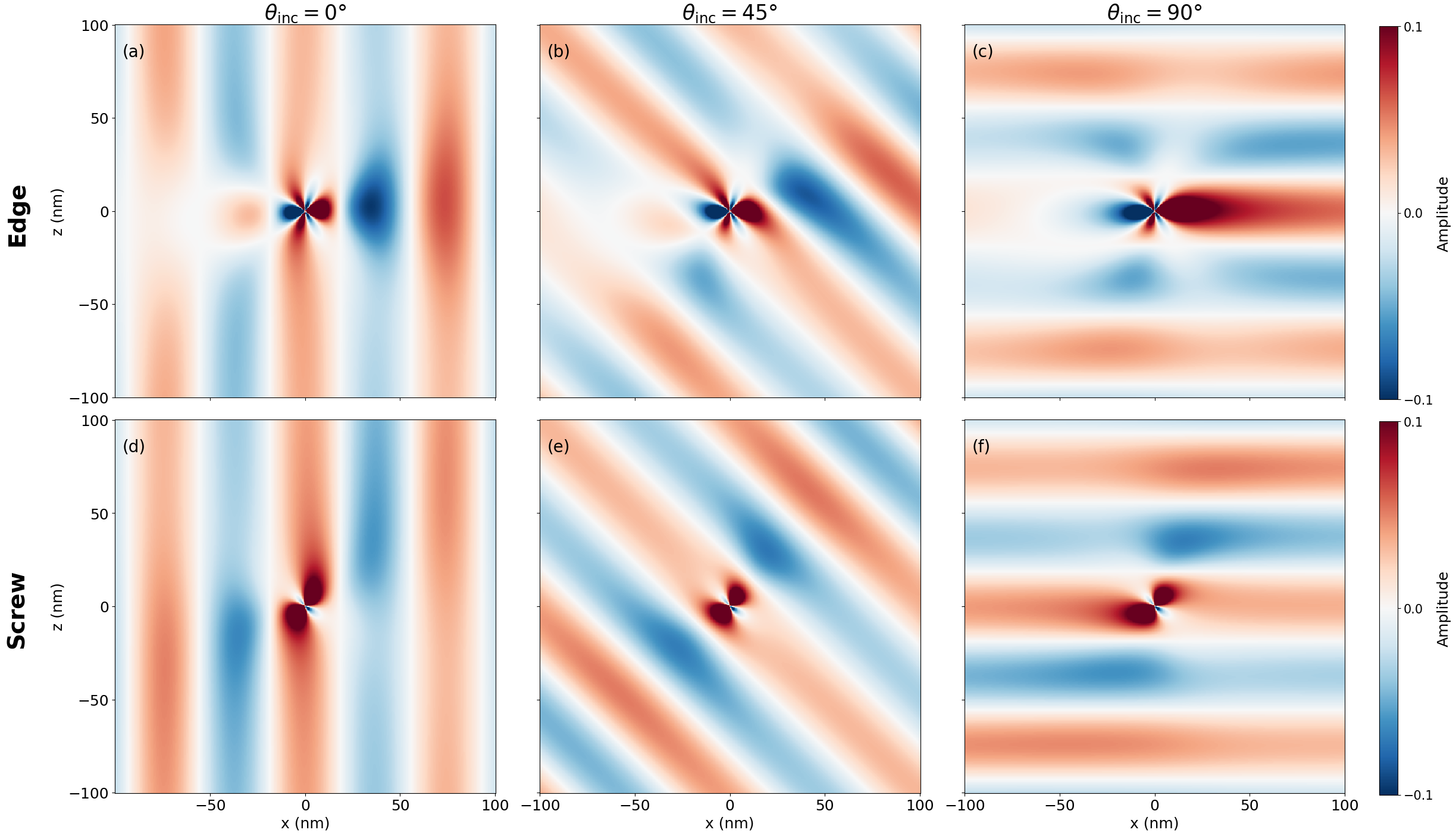}
    \caption{Real part of the scatter wavefunction for the SW in the strong regime. The rows correspond to (top) pure edge and (bottom) pure screw dislocations. The columns display the different incident angles. The incident plane wave propagates along the $+\hat{\bm{x}}$ direction at $\omega/2\pi = 10$ GHz. The color scale indicates normalized amplitude.}
    \label{fig3}
\end{figure*}

\subsection{Two-dimensional scattering}\label{IVA}
We now turn to the full two-dimensional scattering problem. To compute the scattered wave function, we treat the effective potential in Eq. (\ref{schro-like}) as a weak perturbation to $\mathcal{H}_0$	and employ the first Born approximation \cite{cohen}. Within this framework, the scattered wave function is given by the asymptotic expansion
\begin{align}
    \psi({\bs r}) \sim \phi_0({\bs r}) + f(\theta)\frac{e^{ikr}}{\sqrt{r}}, \quad r\rightarrow\infty,
    \label{born}
\end{align}
where $\phi_0({\bs r})$ is the incident plane wave (see Appendix \ref{AppendixC}), and $f(\theta) = \langle{{\bs k}_{\text{sc}}}|V|{{\bs k}_{\text{inc}}}\rangle$ is the scattering amplitude. Here, ${\bs k}_{\ell}=k(\cos\theta_{\ell},\sin\theta_{\ell})$ represents the initial ($\ell=\text{inc}$) and final ($\ell=\text{sc}$) momentum states. The corresponding differential cross-section is therefore given by $d\sigma/d\Omega=|f(\theta)|^2$. In the following, we first examine the real-space SW propagation at a fixed frequency of $\omega/2\pi = 10$ GHz. This frequency is chosen to lie within a propagating band \cite{barman}, thereby avoiding gap regions and enabling a clear visualization of the interference patterns generated by the dislocation line. We then analyze the differential cross-section for different angles of incidence. To emphasize the symmetry signatures imposed by the defect geometry, we focus on the strong-coupling regime. The corresponding analysis in the weak-coupling regime is qualitatively similar but reduced by approximately one order of magnitude (see Appendix \ref{AppendixD}). 

Figure \ref{fig3} displays the real-space dynamics of the total wave function given by Eq. (\ref{born}), for different angles of incidence. For $\theta_{\text{inc}} = 0$, the edge dislocation (Fig. \ref{fig3}(a)) produces a scattering pattern that is symmetric with respect to the defect center, resulting in a balanced bending of the wavefronts around the dislocation core. In contrast, the screw dislocation (Fig. \ref{fig3}(d)) generates an asymmetric response between the upper and lower half-planes along the propagation direction. For $\theta_{\text{inc}} = \pi/4$ and $\theta_{\text{inc}} = \pi/2$, the scattering pattern remains almost unchanged. This demonstrates that the observed differences are not primarily controlled by the incident angle, but rather by the symmetry properties of the effective potential.

\begin{figure}
\centering
\includegraphics[width=0.48\textwidth]{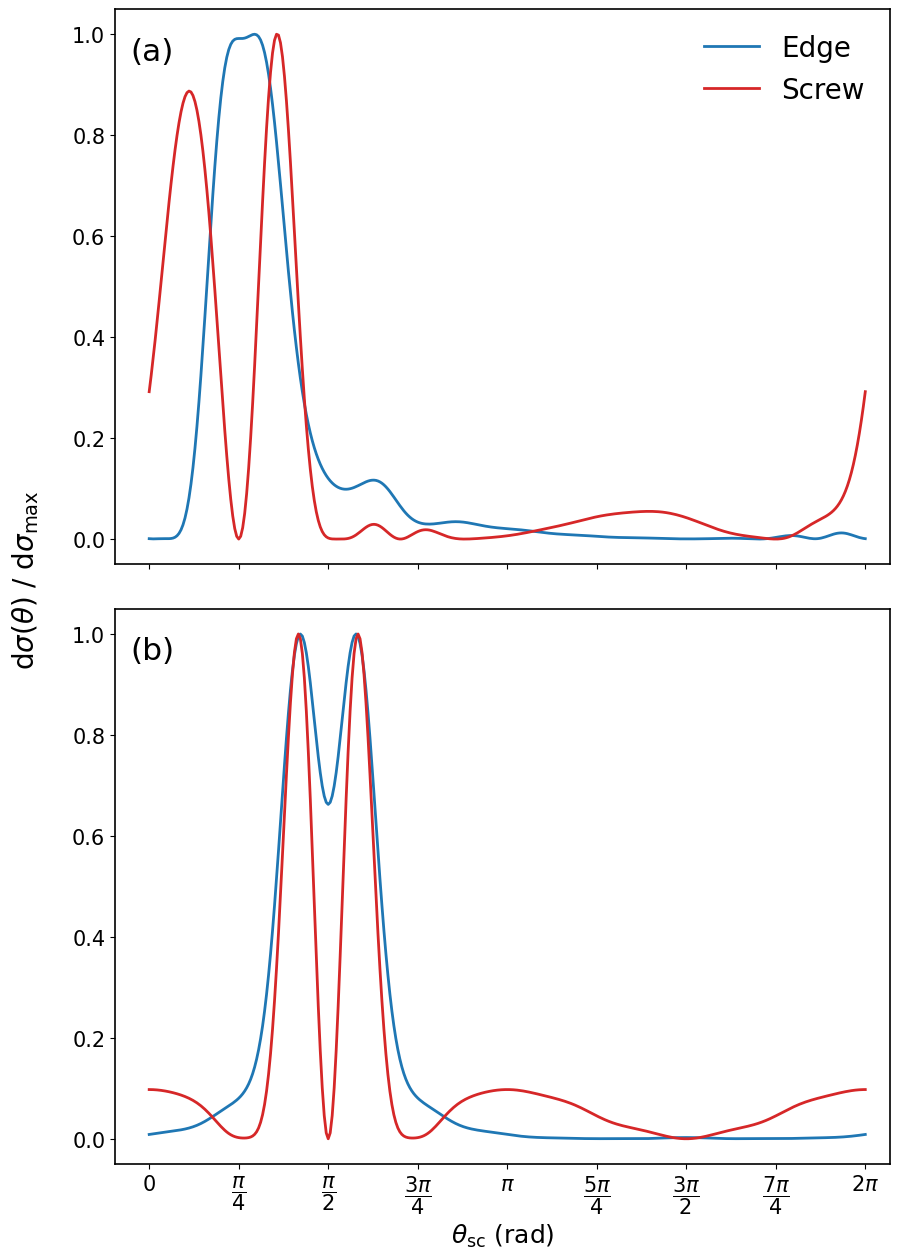}
\caption{Normalized differential cross-section ($d\sigma(\theta_{\text{sc}})/d\sigma_{\text{max}}$) as a function of the scattering angle $\theta_{\text{sc}}$ in the strong regime. (a) $\theta_{\text{inc}} = \pi/4$, and (b) $\theta_{\text{inc}} = \pi/2$. The curve colors distinguish the dislocation type: screw (red) and edge (blue).}
    \label{fig5}
\end{figure}

To quantify these angular distributions, we compute the normalized differential cross-section, where Fig. \ref{fig5} shows the results for two representative incident angles. For an incident angle $\theta_{\text{inc}} = \pi/4$ (Fig. \ref{fig5}(a)), the edge dislocation (blue curve) exhibits a broad single peak centered near the forward direction, indicating that the scattering preserves the symmetry of the incident wave. In contrast, the screw dislocation (red curve) strongly suppresses forward scattering and redistributes the intensity into two lateral lobes, resulting in a qualitatively distinct angular response.
For $\theta_{\text{inc}} = \pi/2$ (Fig. \ref{fig5}(b)), both dislocation types produce angular profiles that are symmetric with respect to the forward axis. However, the edge dislocation retains a predominantly forward-focused distribution, whereas the screw dislocation exhibits a pronounced angular splitting.
\begin{figure*}[!ht]
    \centering
    \includegraphics[width=\textwidth]{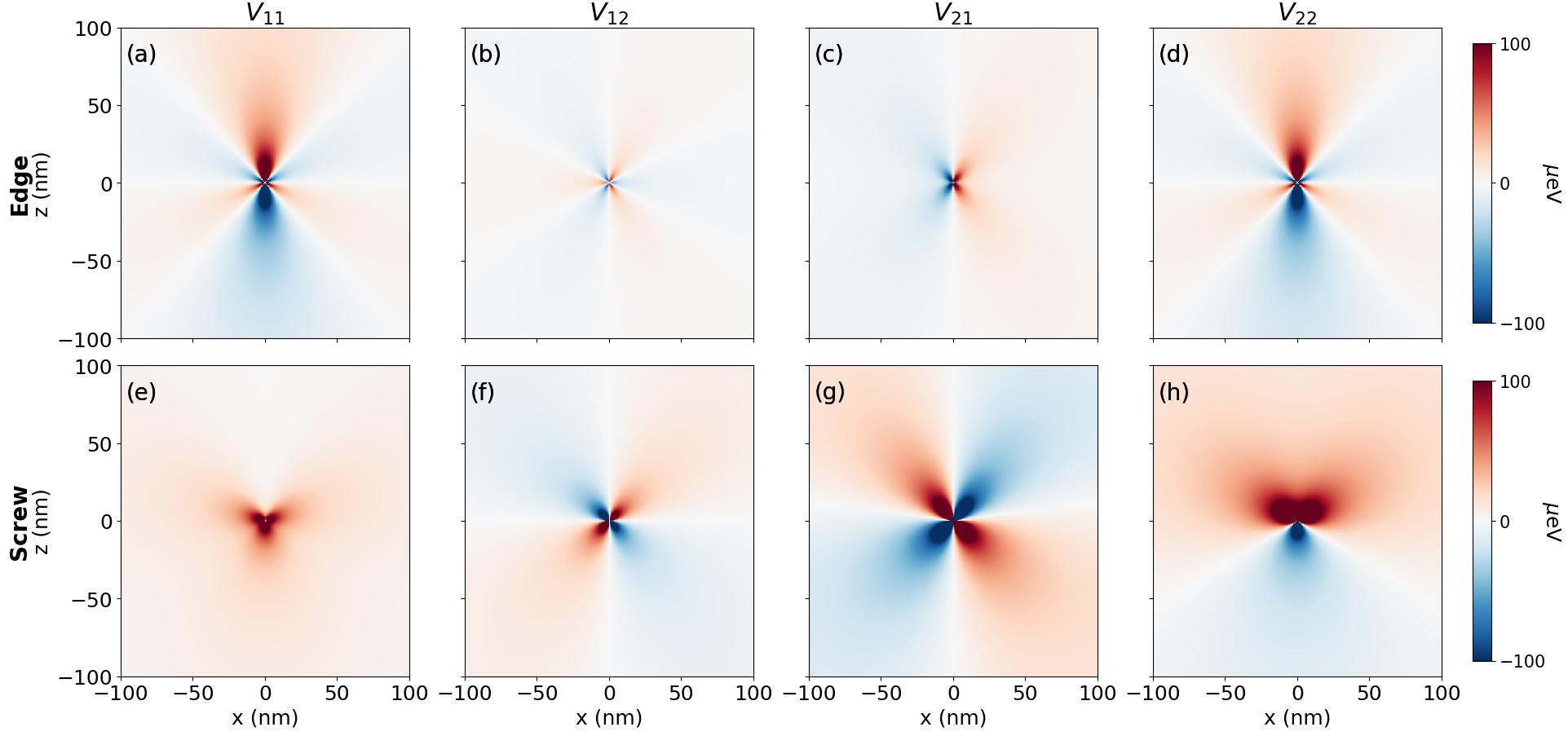}
    \caption{Spatial profiles of the magnetoelastic scattering potential matrix elements $V_{ij}(\mathbf{r})$ in the strong coupling regime. The rows correspond to (top) pure edge and (bottom) pure screw dislocations. The columns display the components of the potential operator. The color scale indicates the potential strength in $\mu$eV.}
    \label{fig4b}
\end{figure*}

Finally, the origin of these distinct scattering signatures lies in the symmetry of the effective potential $V({\bs r})$. Figure \ref{fig4b} shows the matrix elements $V_{ij}$ for edge and screw dislocations in the strong-coupling regime. The scattering response is governed by the interplay between the diagonal terms ($V_{11}, V_{22}$), which act as effective potential barriers, and the off-diagonal terms ($V_{12}, V_{21}$), which induce coupling between the wavefunction components \cite{cohen}. For the edge dislocation (upper row of Fig. \ref{fig4b}), the scattering is dominated by the diagonal terms, which display a pronounced quadrupolar symmetry. Because the off-diagonal terms are negligible (Figs. \ref{fig4b}(b) and \ref{fig4b}(c)), the two wavefunction components remain effectively uncoupled, yielding the symmetric, forward‑focused scattering seen in Figs. \ref{fig3} (top row) and \ref{fig5}. By contrast, for the screw dislocation, the off-diagonal terms are comparable in magnitude to the diagonal contributions (Figs. \ref{fig4b}(f) and \ref{fig4b}(g)). These off-diagonal terms couple the wavefunction components, generating relative phase shifts and producing the asymmetric, split‑lobe angular profiles observed in Figs. \ref{fig3} (bottom row) and \ref{fig5}.

\section{Conclusions and discussion}\label{conclu}
We have investigated the influence of a single straight-line dislocation on the static and dynamic magnetic properties through MEC in a magnetic insulator. We showed that a dislocation line constitutes an active mechanism that generates non-trivial magnetic textures and significantly modifies the SW transport. In the one-dimensional model, we demonstrated that the dislocation line perturbs the homogeneous magnetization locally, generating distortions that depend strictly on the dislocation type. In a domain wall configuration, we demonstrate that the dislocation type can selectively determine the internal wall structure, stabilizing either Néel-type or Bloch-type walls. This structural selectivity could be experimentally tested, offering a novel route for controlling magnetic microstructures through defect engineering. Additionally, the dislocation creates a localized potential barrier that can strongly modulate SW reflection and transmission. For the quasi-homogeneous case, the barrier height can be comparable to or larger than typical SW energies, allowing the system to transition continuously from full transmission to a total reflection. For the domain wall configurations, the presence of the dislocation destroys the intrinsic reflectionless nature and bound states of the Pöschl-Teller potential, transforming it into a repulsive barrier. We showed that the reflectionless behavior is only recovered in the ultra-weak coupling limit, suggesting that real structural defects effectively mask the intrinsic transport properties of domain walls. 

In the two-dimensional system, when the magnetization is initially saturated along the $y$ direction, the dislocation stabilizes a magnetic vortex-like texture localized at the dislocation core. The strength and spatial extent of this texture are controlled by both the dislocation type and the coupling strength. Specifically, a screw dislocation generates stronger in-plane magnetization compared with the mixed dislocation, while an edge dislocation shows minimal coupling. These results demonstrate that the dislocation can act as a source of chiral magnetic textures without the need for Dzyaloshinskii–Moriya interactions. The dislocation-induced magnetic vortex acts as an effective scattering center for SW, exhibiting a highly anisotropic response. Our analysis within the first Born approximation expresses the scattering problem in a form that directly relates the geometry of the defect to the angular structure of the scattered field. This behavior arises from the distinct way each dislocation couples to the magnetization components. The edge dislocation generates a symmetric, diagonal-dominated potential, whereas the screw dislocation introduces significant off-diagonal coupling. Consequently, the system exhibits a preferred scattering symmetry with high-amplitude lobes along the forward direction. These distinct angular scattering signatures could be experimentally verified using spatially resolved techniques, such as micro-focused Brillouin light scattering (BLS) or magneto-optical Kerr effect (MOKE) microscopy, providing a non-invasive method to characterize the local defect landscape.

Finally, this study relies on specific approximations that define its scope. We employed a continuum elasticity framework, which treats the dislocation as a singular line in a continuous medium, and we used the first Born approximation to describe SW scattering. While this perturbative approach captures the essential angular features of the scattering pattern, the amplitudes obtained in the strong coupling regime indicate that higher-order processes may become relevant. In particular, the SW energy $E_{\text{SW}}$ in the absence of dislocations is of the order of 100 GHz ($\approx$ 0.4 meV, see Fig. \ref{disper} in the Appendix), whereas the dislocation-induced potential in the strong regime reaches approximately $V\sim100$ $\mu$eV ($\approx$ 24 GHz). Therefore, the ratio $V/E_{\text{SW}}\approx 0.25$, although still being $<1$, places the system relatively outside the strict perturbative limit required for the quantitative validity of the Born approximation. Consequently, our results should be interpreted as providing a qualitative description of the scattering phenomena, without altering the main physical conclusions regarding symmetry and scattering phenomena. A fully quantitative treatment would benefit from non-perturbative approaches, such as a numerical solution of the Lippmann–Schwinger equation \cite{cohen}. Although real materials typically host complex defect networks rather than the isolated lines modeled here, our single-defect analysis establishes the fundamental physical building blocks for understanding SW propagation in realistic crystalline environments. Ultimately, these findings propose a clear pathway for exploiting structural defects as functional elements in the design of future magnonic devices.

\begin{acknowledgments}
R.E.T, N.V-S, and A.S.N. thank funding from Fondecyt Regular 1230747, 1250364, and 1230515, respectively. This work was partially funded by ANID CEDENNA CIA 250002.
\end{acknowledgments}

\bibliography{Biblio}

\appendix
\begin{widetext}
\section{Hamiltonian components}\label{AppendixA}
We consider small fluctuations of the magnetization around the equilibrium configuration, $\bm{m}(\bm{r},t) = \bm{m}_0(\bm{r}) + \delta\bm{m}(\bm{r},t)$. Accordingly, the effective magnetic field can be decomposed as $\bm{H}_{\text{eff}} = \bm{H}^0(\bm{m}_0) + \delta\bm{H}(\delta\bm{m})$, where $\bm{H}^0$ is the field associated with the ground-state magnetization and $\delta\bm{H}$ is linear in the fluctuations. Neglecting damping, the Landau--Lifshitz--Gilbert equation (Eq. (\ref{LLG})) becomes
\begin{equation}
    \dot{\bm{m}} = \gamma\mu_0 \left[ \bm{m}_0 \times \delta\bm{H} + \delta\bm{m} \times \bm{H}^0 \right].
\end{equation}

This leads to the following system of linearized equations
\begin{align}
    \delta\dot{m}_x =& \frac{2\gamma b_2}{M_s}(m_z\epsilon_{xy} - m_y\epsilon_{xz})\delta m_x \nonumber\\
    &+ \Bigg[\frac{2\gamma A}{M_s}\nabla^2 m_z - \frac{2\gamma A m_z}{M_s}\nabla^2 - \frac{2\gamma K_u m_z}{M_s} - \frac{2\gamma b_1 m_z}{M_s}\epsilon_{zz} - \frac{2\gamma b_2}{M_s}(m_x\epsilon_{xz} + 2m_y\epsilon_{yz})\Bigg]\delta m_y \\
    &+ \Bigg[\frac{2\gamma A m_y}{M_s}\nabla^2 - \frac{2\gamma A}{M_s}\nabla^2 m_y - \frac{2\gamma K_u m_y}{M_s}
    - \frac{2\gamma b_1 m_y}{M_s}\epsilon_{zz} - \frac{2\gamma b_2}{M_s}(m_x\epsilon_{xy} + 2m_z\epsilon_{yz})\Bigg]\delta m_z ,\nonumber
\end{align}

\begin{align}
    \delta\dot{m}_y =& \Bigg[\frac{2\gamma A m_z}{M_s}\nabla^2 - \frac{2\gamma A}{M_s}\nabla^2 m_z  + \frac{2\gamma b_1 m_z}{M_s}(\epsilon_{zz}-\epsilon_{xx}) + \frac{2\gamma b_2}{M_s}(m_y\epsilon_{yz} - 2m_x\epsilon_{xz})\Bigg]\delta m_x \nonumber\\
    &+ \frac{2\gamma b_2}{M_s}(m_x\epsilon_{yz} - m_z\epsilon_{xy})\delta m_y \\
    &+ \Bigg[\frac{2\gamma A}{M_s}\nabla^2 m_x - \frac{2\gamma A m_x}{M_s}\nabla^2
    + \frac{2\gamma b_1 m_x}{M_s}(\epsilon_{zz}-\epsilon_{xx}) - \frac{2\gamma b_2}{M_s}(m_y\epsilon_{xy} + 2m_z\epsilon_{xz})\Bigg]\delta m_z ,\nonumber
\end{align}

\begin{align}
    \delta\dot{m}_z =& \Bigg[\frac{2\gamma A}{M_s}\nabla^2 m_y - \frac{2\gamma A m_y}{M_s}\nabla^2
    + \frac{2\gamma K_u m_y}{M_s} + \frac{2\gamma b_1 m_y}{M_s}\epsilon_{xx} - \frac{2\gamma b_2}{M_s}(m_z\epsilon_{yz} + 2m_x\epsilon_{xy})\Bigg]\delta m_x \nonumber\\
    &+ \Bigg[\frac{2\gamma A m_x}{M_s}\nabla^2 - \frac{2\gamma A}{M_s}\nabla^2 m_x
    + \frac{2\gamma K_u m_x}{M_s} + \frac{2\gamma b_1 m_x}{M_s}\epsilon_{xx} + \frac{2\gamma b_2}{M_s}(m_z\epsilon_{xz} + 2m_y\epsilon_{xy})\Bigg]\delta m_y\\
    &+ \frac{2\gamma b_2}{M_s}(m_y\epsilon_{xz} - m_x\epsilon_{yz})\delta m_z .\nonumber
\end{align}
where the strain tensor $\epsilon_{ij}$ is given by
\begin{equation}
    \epsilon = \left(\begin{array}{ccc}
        -\dfrac{b_x}{4\pi}\dfrac{(3-2\nu)x^2z + (1-2\nu)z^3}{(1-\nu)(x^2+z^2)^2} & -\dfrac{b_y}{4\pi}\dfrac{z}{x^2+z^2} & \dfrac{b_x}{4\pi}\dfrac{(z^2 - x^2)x}{(1-\nu)(x^2+z^2)^2}\\[10pt] 
        -\dfrac{b_y}{4\pi}\dfrac{z}{x^2+z^2} & 0 & \dfrac{b_y}{4\pi}\dfrac{x}{x^2+z^2} \\[10pt] 
        \dfrac{b_x}{4\pi}\dfrac{(z^2 - x^2)x}{(1-\nu)(x^2+z^2)^2} & \dfrac{b_y}{4\pi}\dfrac{x}{x^2+z^2} & \dfrac{b_x}{4\pi}\dfrac{(1+2\nu)x^2z - (1-2\nu)z^3}{(1-\nu)(x^2+z^2)^2}
    \end{array}\right).
    \label{strain}
\end{equation}
This system can be written compactly as $\delta\dot{\bm{m}} = M\,\delta\bm{m}$, where $M$ is a $3\times3$ matrix. To describe SW dynamics, we project the system onto a local orthonormal basis perpendicular to the equilibrium magnetization. We define
\begin{equation}
    \hat{\bm{r}} = \left(\begin{array}{c}
        m_x\\
        m_y\\
        m_z    
    \end{array}\right), \quad \hat{\bm{e}}_1 = \frac{1}{\sqrt{m_x^2+m_y^2}}\left(\begin{array}{c}
        m_y\\
        -m_x\\
        0   
    \end{array}\right), \quad \hat{\bm{e}}_2 = \frac{1}{\sqrt{m_x^2+m_y^2}}\left(\begin{array}{c}
        m_xm_z\\
        m_ym_z\\
        -(m^2_x + m_y^2)
    \end{array}\right)
\end{equation}
The corresponding rotation matrix is $\mathcal{R} = (\hat{\bm{r}}\, \hat{\bm{e}}_1\, \hat{\bm{e}}_2)$. Introducing the transverse fluctuation vector $\psi =(0,\delta m_1,\delta m_2)^T$ and projecting onto the orthonormal basis, this yields $\dot{\psi} =(\mathcal{H}_0 + V)\psi$ where
\begin{equation}
    \mathcal{H}_0 =
    \begin{pmatrix}
        \alpha_0 & \beta_0 \\
        \gamma_0 & \lambda_0
    \end{pmatrix}\qquad \text{and}\qquad V =
    \begin{pmatrix}
        V_{11} & V_{12} \\
        V_{21} & V_{22}
    \end{pmatrix}.
\end{equation}
The explicit expressions for the free Hamiltonian are,
\begin{align}
    \alpha_0 =& \frac{2\gamma K_u m_x m_y m_z}{M_s\rho^2} + \frac{2\gamma A m_z}{M_s} \bigl(\nabla^2 \varphi + 2 \nabla \varphi \cdot \nabla\bigr) \\[0.5ex]
    \beta_0 =& -\frac{2\gamma A}{M_s}\nabla^2 + \frac{2\gamma K_u m_y^2}{M_s\rho^2} (m_x^2+m_y^2-m_z^2) +\frac{2\gamma A}{M_s}\left[m_y\nabla^2 m_y - m_x\nabla^2 m_x + \frac{m_y^2 - m_x^2}{m_x^2 + m_y^2} m_z\nabla^2 m_z \right] \nonumber\\
    &\quad - \frac{2\gamma A}{M_s}\left[\rho\nabla^2\rho + m_z\nabla^2 m_z - m_z^2(\nabla\varphi)^2 + 2(\rho\nabla\rho + m_z\nabla m_z)\cdot\nabla \right]\\
    \gamma_0 =& \frac{2\gamma A}{M_s}\nabla^2 + \frac{2\gamma K_u}{M_s\rho^2}(m_x^4-m_y^4+m_x^2m_z^2) - \frac{2\gamma A}{M_s}\left[m_y\nabla^2m_y - m_x\nabla^2m_x + m_z\nabla^2m_z\right] - \frac{2\gamma A}{M_s}(\nabla\varphi)^2\\
    \lambda_0 =& -\frac{2\gamma K_u m_xm_ym_z}{M_s\rho^2} - \frac{4\gamma Am_ym_z}{M_s\rho^2}[(m_x^2+m_y^2)\nabla^2m_x + m_xm_z\nabla^2m_z] \nonumber\\
        &+ \frac{2\gamma A}{M_s}(m_z\nabla^2\varphi + 2\nabla m_z\nabla\varphi + 2m_z\nabla\varphi\cdot\nabla)
\end{align}
where $\rho \equiv \sqrt{m^2_x + m^2_y}$, $\varphi \equiv \arctan(m_y/m_x)$. For the magnetoelastic interaction potential
\begin{align}
     V_{11} =& \frac{2\gamma b_1 m_xm_ym_z\epsilon_{xx}}{M_s\rho^2} + \frac{2\gamma b_2m_y}{M_s\rho^2}\left[(m_x+m_y)m_z\epsilon_{xy} + (m_y-m_x)m_z\epsilon_{zy} - (m^2_x + m^2_y)\epsilon_{xz}\right] \\
     V_{12} =& \frac{2\gamma b_1(m_x^2+m_y^2-m_z^2)}{M_s\rho^2}\left[m_x^2\epsilon_{xx} + (m_y^2-m_x^2)\epsilon_{zz}\right]\nonumber\\
        &+ \frac{2\gamma b_2m_z}{M_s\rho^2}\left[(m_y+m_x)m_xm_z\epsilon_{xy} + 4m_x^3\epsilon_{xz} + (m^3_x - m_x^2m_y - 4m^3_y)\epsilon_{zy}\right]\\
    V_{21} =& \frac{2\gamma b_1}{M_s\rho^2}\left[(m_x^4 - m_y^4 - m_y^2m_z^2)\epsilon_{xx} + (m_x^2+m_y^2)m_z^2\epsilon_{zz}\right] \nonumber\\
        &+ \frac{2\gamma b_2}{M_s\rho^2}\Big\{\left[4m_x(m_x^2+m_y^2) + (m_x-m_y)m_z^2\right]m_y\epsilon_{xy} \\
        &\hspace{5cm} + 2m_xm_z(m_x^2+m_y^2)\epsilon_{xz} + (3m_x^3 + m_xm_y + 2m_y^2)m_zm_y\epsilon_{zy}\Big\}\nonumber\\
    V_{22} =& \frac{2\gamma b_1 m_xm_ym_z\epsilon_{xx}}{M_s\rho^2}[(m_z^2 - 3m_x^2 - 3m_y^2)\epsilon_{xx} + 2(m_x^2+m_y^2-m_z^2)\epsilon_{zz}] \nonumber\\
        & + \frac{2\gamma b_2}{M_s\rho^2}\Big\{[-m_x^4 + 2m_x^2m_y^2 + 3m_y^4 - m_xm_z^2(m_x+m_y)]m_z\epsilon_{xy} \\
        &\hspace{5cm} + [-m_x^4 - 2m_x^2m_y^2 - m_y^4 + m_z^2(7m_x^2+3m_y^2)]m_y\epsilon_{xz} \nonumber\\
        &\hspace{5cm} + [m_x^4 + 2m_x^2m_y^2 + m_y^4 + m_x^2m_z^2 + m_ym_z^2(m_x+4m_y)]\epsilon_{zy}\Big\}\nonumber
\end{align}

\subsection{One-dimensional approximation}\label{AppendixB}
At $z=0$, the strain tensor is reduced to
\begin{equation}
    \epsilon|_{z=0} = \left(\begin{array}{ccc}
        0 & 0 & \dfrac{b_x}{4\pi x(1-\nu)}\\
        0 & 0 & \dfrac{b_y}{4\pi x} \\
        \dfrac{b_x}{4\pi x(1-\nu)} & \dfrac{b_y}{4\pi x} & 0
    \end{array}\right).\label{straintensor}
\end{equation}

Figure \ref{fig_GS1DW} shows the different ground state configurations for the homogeneous and DW conditions in the strong regime. Figure \ref{potentialxRT_strong}a shows the effective potential for each dislocation type in both coupling regimes for the quasi-homogeneous case, and Fig. \ref{potentialxRT_strong}b shows the reflection and transmission coefficients for the strong regime in the quasi-homogeneous case. Figure \ref{potentialxRT_DW}a shows the effective potential for each dislocation type in both coupling regimes for the domain wall case, and Fig. \ref{potentialxRT_DW}b shows the reflection and transmission coefficients for the weak regime in the domain wall case.
\begin{figure}[!h]
    \centering
    \includegraphics[width=.8\textwidth]{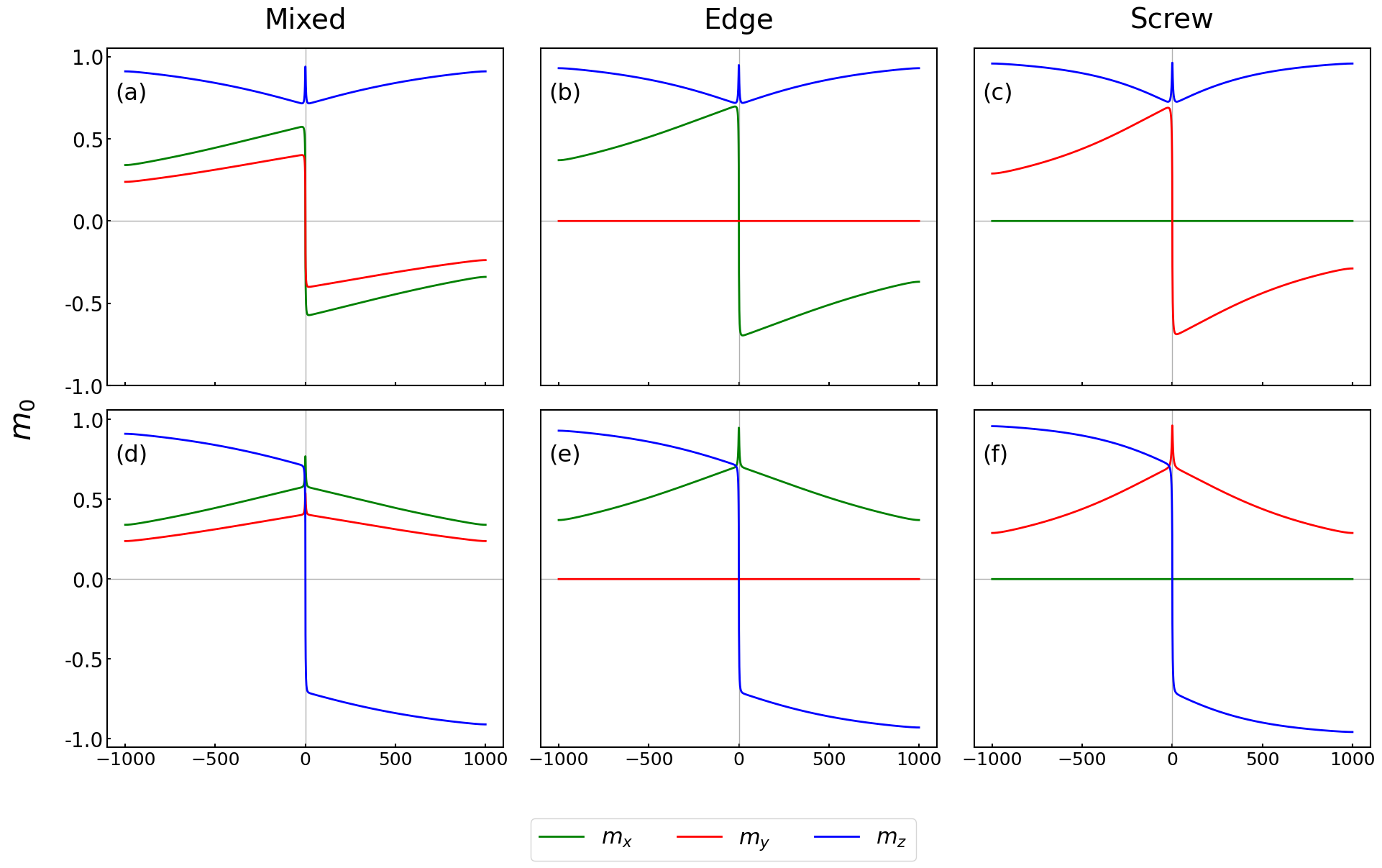}
    \caption{Relaxed magnetization profiles for the one-dimensional model in the strong regime. Panels (a)-(c) show the relaxed profiles for the homogeneous boundary conditions, and panels (d)-(f) show the relaxed profiles for the domain wall conditions. Columns correspond to mixed, edge, and screw dislocation types, respectively.}
    \label{fig_GS1DW}
\end{figure}
\begin{figure}[!h]
    \centering
    \includegraphics[width=\textwidth]{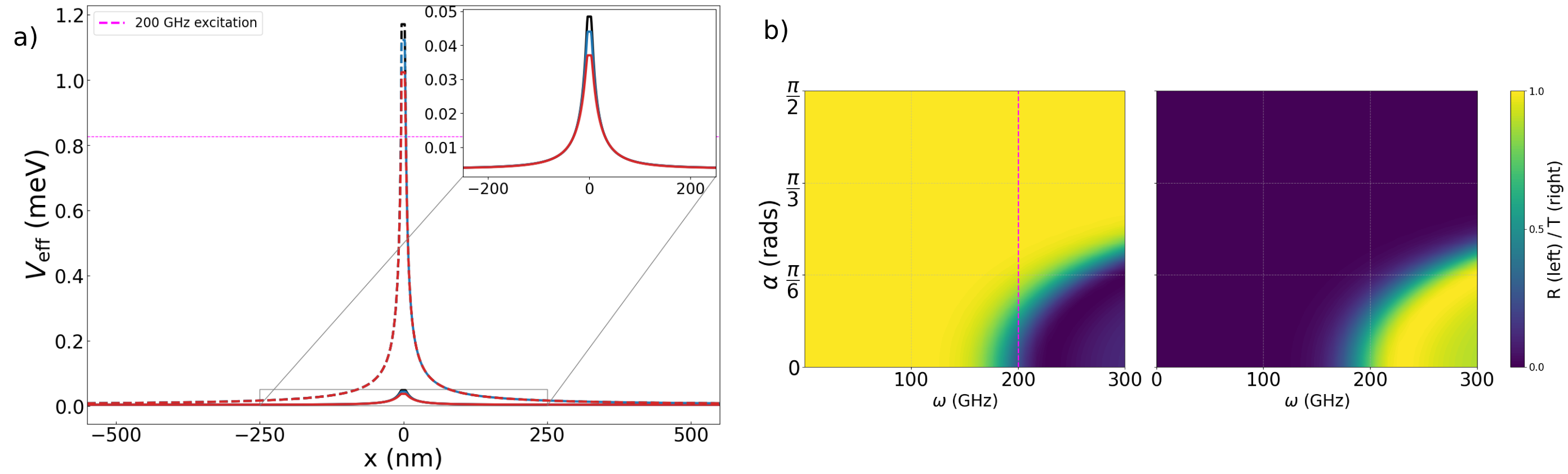}
    \caption{(a) One-dimensional effective potential profiles for different dislocation types in the quasi-homogeneous case for the strong regime. The inset provides a magnified view of the weak coupling regime (solid lines) near the origin. (b) Color maps showing the intensity of reflection ($R$) and transmission ($T$) for a mixed dislocation across a frequency range of 0–300 GHz and incident angles $\alpha$ from 0 to $\pi/2$. The dashed vertical line indicates the 200 GHz excitation reference.}
    \label{potentialxRT_strong}
\end{figure}
\begin{figure}[!h]
    \centering
    \includegraphics[width=\textwidth]{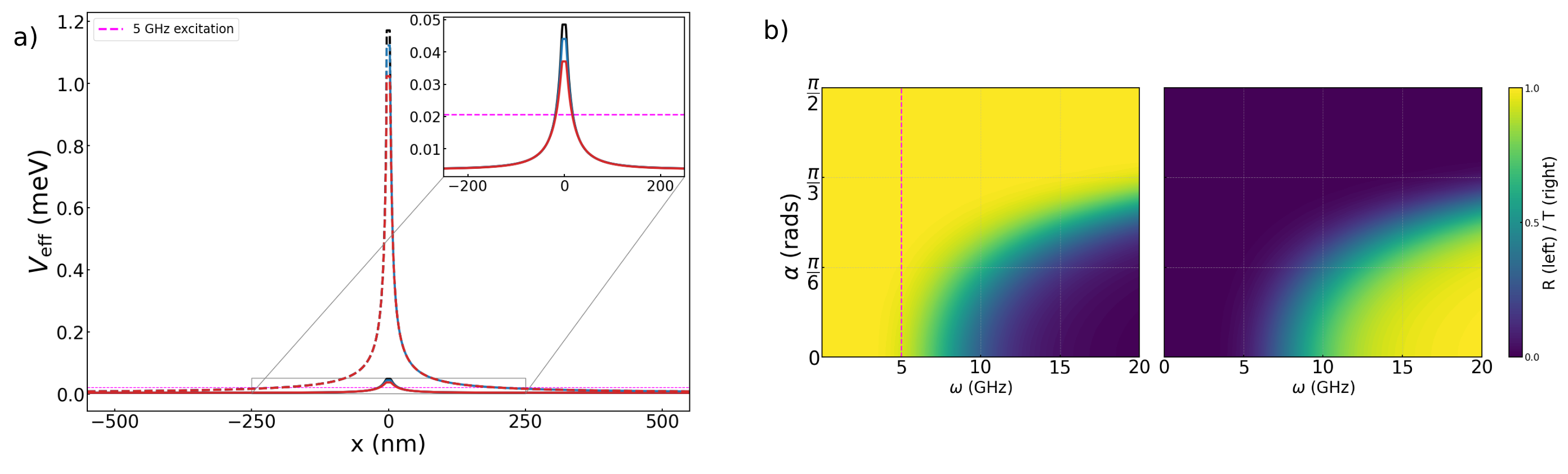}
    \caption{(a) One-dimensional effective potential profiles for different dislocation types in the DW case. The inset provides a magnified view of the weak coupling regime (solid lines) near the origin. (b) Color maps showing the intensity of reflection ($R$) and transmission ($T$) for a mixed dislocation across a frequency range of 0–20 GHz and incident angles $\alpha$ from 0 to $\pi/2$. The dashed vertical line indicates the 5 GHz excitation reference.}
    \label{potentialxRT_DW}
\end{figure}

\newpage
\subsection{Incident wave}\label{AppendixC}
To obtain the incident wave for the scattering in the first Born approximation, we need to solve the eigenvalue problem
\begin{equation}
    \left(\begin{array}{cc}
        \tilde{\alpha}_0 & \tilde{\beta}_0  \\
        \tilde{\gamma}_0 & \tilde{\lambda}_0
    \end{array}\right)\left(\begin{array}{c}
        u \\
        v
    \end{array}\right) = \omega_{+}\left(\begin{array}{c}
        u \\
        v
    \end{array}\right) ,
\end{equation}
where $u$ and $v$ are the eigenvectors to solve, and $\omega_{+}$ is given by Eq. (\ref{omegapm}). This system of equations yields
\begin{equation}
    v = \frac{\omega_{+} - \tilde{\alpha}_0}{\tilde{\beta}_0} \quad\Rightarrow\quad \phi_0 = \left(\begin{array}{c}
        1 \\
        \dfrac{\omega_{+} - \tilde{\alpha}_0}{\tilde{\beta}_0}
    \end{array}\right)
\end{equation}
Since we are dealing with SW, the normalization is given by $|u|^2 + |v|^2 = 1$. Therefore, the incident wave is given by 
\begin{equation}
    \phi_0(x,z) = \frac{1}{\sqrt{1 + \left|\frac{\omega - \alpha_0}{\beta_0}\right|}}\left(\begin{array}{c}
                    1 \\
                    \dfrac{\omega - \alpha_0}{\beta_0} 
                \end{array}\right)
\end{equation}

\subsection{Two-dimensional Scattering in the weak regime}\label{AppendixD}
Figure \ref{fig3_weak} shows the spatial dynamics of the total wavefunction in the weak regime, where different columns represent different incident angles. Figure \ref{fig4b} illustrates the matrix elements $V_{ij}$ for edge and screw dislocations in the weak regime.
\begin{figure}[!h]
    \centering
    \includegraphics[width=0.8\textwidth]{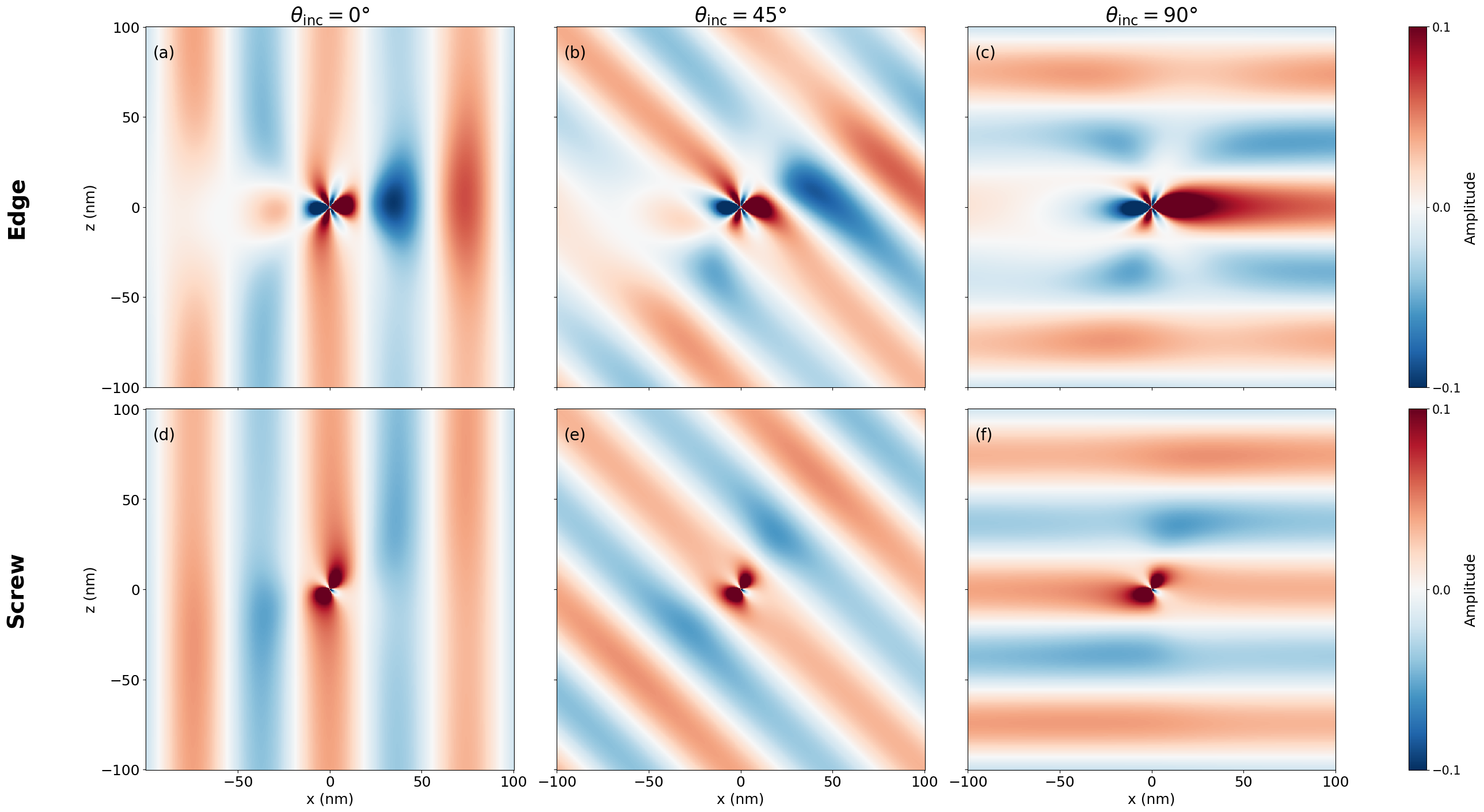}
    \caption{Real part of the scatter wavefunction for the SW in the weak regime. The rows correspond to (top) pure edge and (bottom) pure screw dislocations. The columns display the different incident angles. The incident plane wave propagates along the $+\hat{\bm{x}}$ direction at $\omega/2\pi = 10$ GHz. The color scale indicates normalized amplitude.}
    \label{fig3_weak}
\end{figure}
\begin{figure}[!h]
    \centering
    \includegraphics[width=\textwidth]{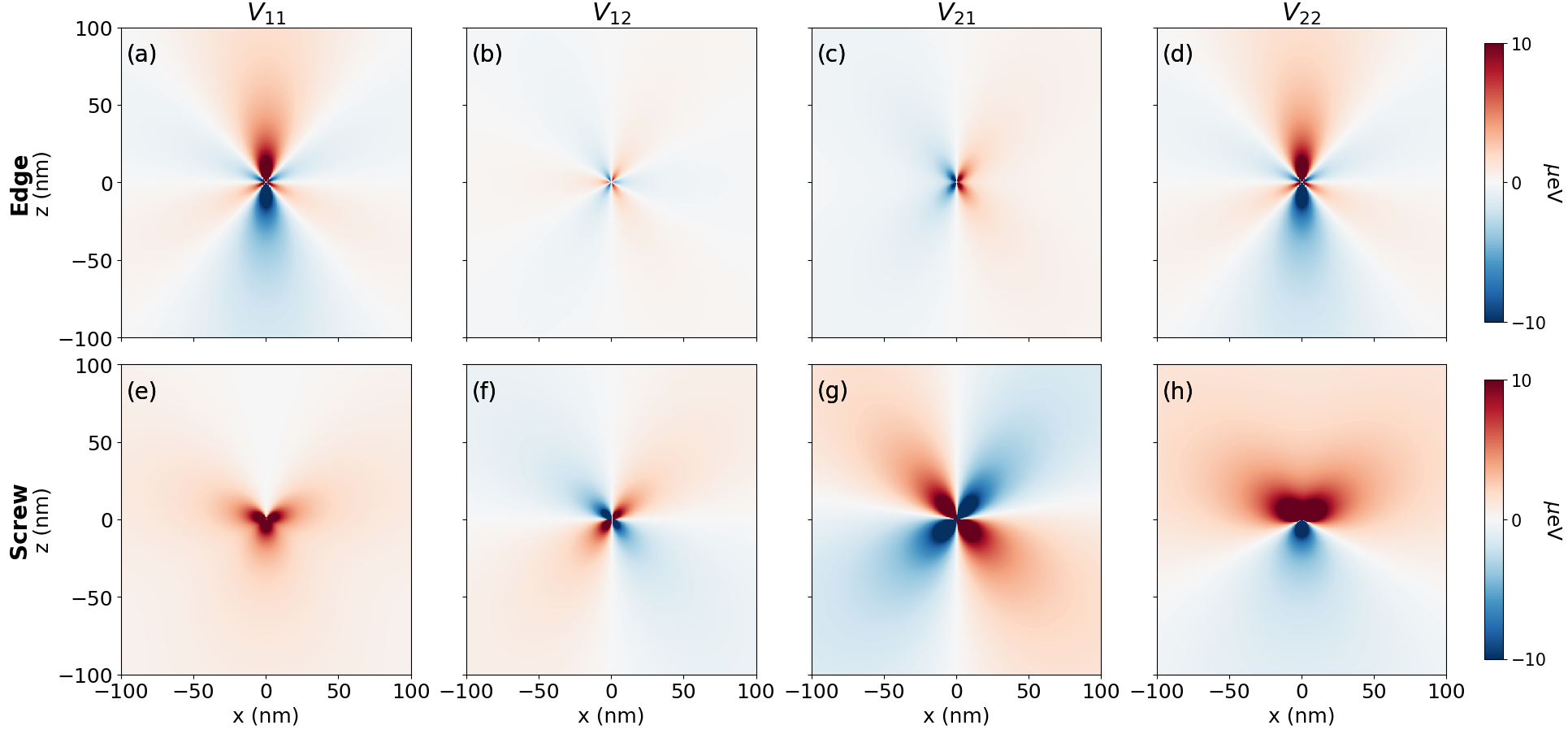}
    \caption{Spatial profiles of the magnetoelastic scattering potential matrix elements $V_{ij}(\mathbf{r})$ in the weak coupling regime. The rows correspond to (top) pure edge and (bottom) pure screw dislocations. The columns display the components of the potential operator. The color scale indicates the potential strength in $\mu$eV.}
    \label{fig4b_weak}
\end{figure}

\subsection{Dispersion Relation}\label{AppendixE}
To obtain the complete expression for the dispersion relation, we define $\bm{k} = k(\cos a, \sin a)$ and we keep the definitions for $\rho$ and $\varphi$. Also, we define 
\begin{align}
    A({\bs r}) &\equiv \frac{2\gamma K_u m_xm_ym_z}{M_s\rho^2} + \frac{2\gamma Am_z}{M_s}\nabla^2\varphi\\
    \tilde{A} &\equiv \frac{2\gamma A}{M_s}\\
    B({\bs r}) &\equiv \frac{4\gamma Am_z}{M_s}((\nabla\varphi)_x\cos a + (\nabla\varphi)_z\sin a) \\
    \begin{split}
        C({\bs r}) &\equiv \frac{2\gamma K_u m^2_y (m_x^2+m_y^2-m_z^2)}{M_s\rho^2} + \frac{2\gamma A}{M_s}\left[m_y\nabla^2m_y - m_x\nabla^2m_x + \frac{m_y^2 - m_x^2}{m_x^2 + m_y^2} m_z\nabla^2m_z\right]\\
        &\hspace{5cm}- \frac{2\gamma A}{M_s}\left[\rho\nabla^2\rho + m_z\nabla^2m_z - m_z^2(\nabla\varphi)^2\right]
    \end{split}\\
    D({\bs r}) &\equiv \frac{4\gamma A}{M_s}((\rho\nabla\rho + m_z\nabla m_z)_x \cos a + (\rho\nabla\rho + m_z\nabla m_z)_z \sin a)\\
    E({\bs r}) &\equiv \frac{2\gamma K_u}{M_s\rho^2}(m_x^4-m_y^4+m_x^2m_z^2) - \frac{2\gamma A}{M_s}\left[m_y\nabla^2m_y - m_x\nabla^2m_x + m_z\nabla^2m_z\right] - \frac{2\gamma A}{M_s}(\nabla\varphi)^2\\
    F({\bs r}) &\equiv -\frac{2\gamma K_u m_xm_ym_z}{M_s\rho^2} - \frac{4\gamma Am_ym_z}{M_s\rho^2}[(m_x^2+m_y^2)\nabla^2m_x + m_xm_z\nabla^2m_z] + \frac{2\gamma A}{M_s}(m_z\nabla^2\varphi + 2\nabla m_z\nabla\varphi).
\end{align}

Then, the free Hamiltonian can be written as
\begin{equation}
    \mathcal{H}_0 = \left(\begin{array}{cc}
        A + iBk & C - iDk + \tilde{A}k^2  \\
        E-\tilde{A}k^2 & F + iBk
    \end{array}\right)
\end{equation}
and the dispersion relation obeys the equation
\begin{align}
    &\det(-i\omega - \tilde{\mathcal{H}}_0) = 0\\
    &\left|\begin{array}{cc}
        -i\omega -A - iBk & C - iDk + \tilde{A}k^2  \\
        E-\tilde{A}k^2 & -i\omega - F - iBk
    \end{array}\right| = 0\\
    \Rightarrow \quad&\omega^2 + 2\omega Bk + B^2k^2 - i(\omega + Bk)(A+F) + CE - iDEk - \tilde{A}Ck^2 + \tilde{A}Ek^2 +i\tilde{A}Dk^3 - \tilde{A}^2k^4 =0
\end{align}

Taking the real part of this equation
\begin{equation}
    \omega_{\pm} = - B({\bs r})k \pm \sqrt{\tilde{A}^2k^4 + \tilde{A}(C({\bs r})-E({\bs r}))k^2 - C({\bs r})E({\bs r})}.
    \label{omegapm}
\end{equation}
Figure \ref{disper} shows the dispersion relation in the bulk.
\begin{figure}[!h]
    \centering
    \includegraphics[width=0.55\textwidth]{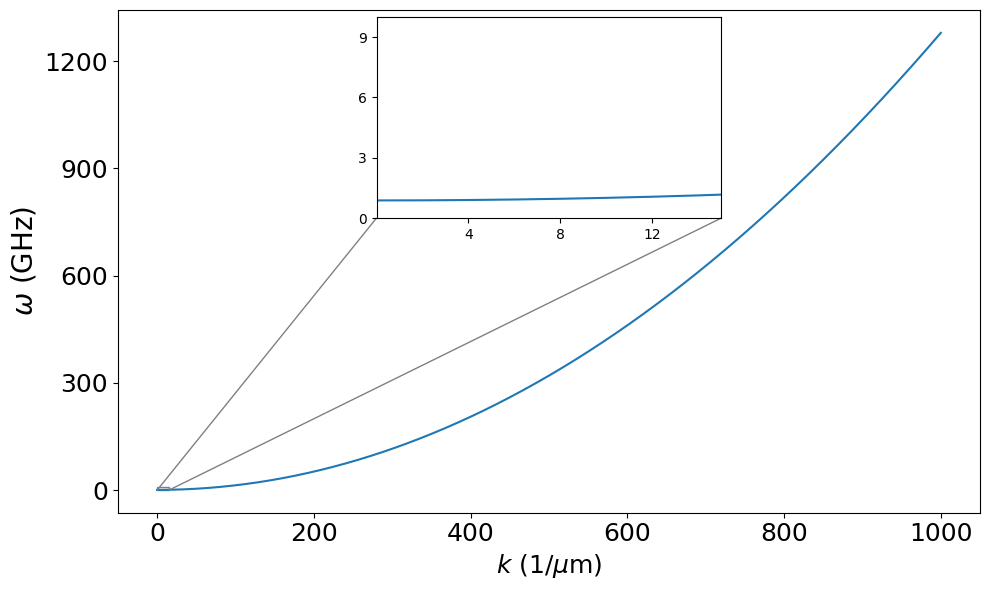}
    \caption{Analytical dispersion relation $\omega(k)$ for bulk SW. The main plot illustrates the quadratic behavior dominated by the exchange interaction at high wave vectors. Inset: Detail of the long-wavelength limit ($k \to 0$), showing the finite frequency gap induced by magnetic anisotropy.}
    \label{disper}
\end{figure}

\end{widetext}

\end{document}